\documentclass[epj,final]{svjour}
\usepackage{latexsym}
\usepackage{graphics}
\usepackage{amsmath,amssymb}

\newcommand{\PsilrO}{\chi_{\lambda}(r,\Omega)}

\newcommand{\Nln}{{\cal N}_{nl} }
\newcommand{\jlkr}{j_l(\kappa_{nl}r) }
\newcommand{\cYlmnO}{{\cal Y}_{l}^{m}(\Omega) }

\newcommand{\Cllbar}{\overline{C}_{\lambda \lambda'}}
\newcommand{\dll}{\delta_{\lambda \lambda'}}
\newcommand{\CllV}{C_{\lambda \lambda'}^{(V)} }
\newcommand{\Cllt}{C_{\lambda \lambda'}^{(\tau)} }
\newcommand{\Clltbar}{\overline{C}_{\lambda \lambda'}^{(\tau)} }

\newcommand{\CllW}{C_{\lambda \lambda'}^{(W)} }
\newcommand{\CllS}{C_{\lambda \lambda'}^{(S)} }
\newcommand{\CllC}{C_{\lambda \lambda'}^{(C)} }
\newcommand{\vgrad}{\vec{\nabla}}
\newcommand{\ql}{q_{\lambda}}
\newcommand{\qlp}{q_{\lambda'}}
\newcommand{\rmd}{\mathrm{d}}
\newcommand{\rme}{\mbox{e}}
\begin{document}
\title{Stability and instability of a hot and dilute nuclear droplet}
\subtitle{I. Adiabatic isoscalar modes}
\author{W.~N\"orenberg\inst{1,2} 
\and G.~Papp\inst{1,3,}\thanks{\emph{Present address:} 
HAS Research Group for Theoretical Physics, E\"{o}tv\"{o}s University,
Budapest, P\'azm\'any P. s. 1/A, H-1117 Budapest, Hungary}
\and P.~Rozmej\inst{1,4,}\thanks{\emph{Present address:} 
 Institute of Mathematics, Technical University of Zielona G\'ora, 
 ul.\ Podg\'orna 50, Pl-65246 Zielona G\'ora, Poland
\protect\newline 
\hspace*{4ex}\emph{e-mail addresses:} 
w.nrnbrg@gsi.de, pg@hal9000.elte.hu, p.rozmej@im.pz.zgora.pl}%
}                     
%
\institute{Gesellschaft f\"ur Schwerionenforschung, 
D-64291 Darmstadt, Germany \and 
Institut f\"ur Kernphysik, Technische Universit\"at  Darmstadt,
D-64289 Darmstadt, Germany \and
Institut f\"ur Theoretische Physik, Universit\"at Heidelberg, 
D-69120 Heidelberg, Germany
\and Instytut Fizyki, Uniwersytet Marii Curie-Sk\l odowskiej, 
Pl-20031 Lublin, Poland 
}
\date{Received: \today }
%
\abstract{
The diabatic approach to collective nuclear motion is reformulated in the
local-density appro\-ximation in order to treat the normal modes of a
spherical nuclear droplet analytically. In a first application 
the adiabatic
isoscalar modes are studied and results for the eigenvalues of
compressional (bulk) and pure surface modes are presented as function of
density and temperature inside the droplet,
as well as for different mass numbers and for
soft and stiff equations of state. We find that the region of bulk 
instabilities (spinodal regime) is substantially smaller 
for nuclear droplets than for infinite nuclear matter. 
For small densities below
30\% of normal nuclear matter density and for temperatures below 5 MeV
all relevant bulk modes become unstable with the same growth rates.
The surface modes have a larger spinodal region, reaching out to
densities and temperatures way beyond the spinodal line for bulk
instabilities. Essential experimental features of multifragmentation, 
like fragmentation temperatures and
fragment-mass distributions (in particular the power-law behavior)
are consistent with the instability properties of an expanding
nuclear droplet, and hence with a dynamical fragmentation
process within the spinodal regime of bulk and surface modes (spinodal
decomposition).
} 
\maketitle
\section{Introduction and summary}\label{intro}

In bombardments of nuclei with light and heavy ions, as well as
by absorption of pions and antiprotons in nuclei, transient systems
are produced at excitation energies around 10 MeV/u, which decay 
into several fragments of intermediate size.
Increasing interest \cite{b1,b1*,b2,b3} in this 
multifragmentation reaction has been initiated by the observation 
\cite{b4} that the fragment-mass distribution is proportional to
$A^{-\sigma}$ (power law with $\sigma \approx 2.6$) indicating that
the process may be related to the critical point of the liquid-gas 
phase transition of nuclear matter. Indeed, simulations within 
molecular dynamics \cite{b7,b6,b5} and mean-field 
approaches \cite{b8,b9,b10,b11} show that initial
compression in central heavy-ion collisions or pure excitation of a 
nucleus by abrasion of nucleons or by absorption of light ions, pions
or antiprotons cause the system to expand and to break up into pieces
in a low-density regime. Power-law fragment-mass distributions have
also been obtained from dynamical nucleation in a thermodynamically
unstable nucleonic system \cite{b11*}.
However, as indicated by the great success
of statistical models \cite{b12,b13,b14,b15},
the observables in multifragmentation seem to be mainly determined 
by the available phase space at the point of freeze out \cite{b17*}. 
This conclusion remains valid even if the time evolution (expansion) 
of the statistically emitting source is taken into account 
\cite{b16}.

With the discovery of a caloric curve in projectile fragmentation
by the ALADIN collaboration \cite{b17}, the nuclear liquid-gas phase
transition as a mediator between the heat\-ed expanding nuclear matter 
and its decay into pieces moved again into the center of discussions.
The growth of density instabilities within the spinodal 
region inside the phase coexisting regime, known as spinodal 
decomposition \cite{b18}, has early been considered as a possible 
mechanism of multifragmentation \cite{b19,b20,b21}. 
Extending Landau's Fermi-liquid theory to finite
temperatures Heiselberg, Pet\-hick and Ravenhall \cite{b22} have
investigated the stability and instability of hot and dilute 
infinite nuclear matter. 
Subsequent studies have been concerned with detailed properties
of nuclear matter \cite{b23} and finite droplets 
\cite{b26,b25,b27,b24} in and around the spinodal regime.
Quantal effects, as studied within constrained RPA \cite{b25}, 
do not alter the adiabatic instabilities as calculated from a 
fluid-dynamical description \cite{b26} and simulations \cite{b27}
within a stochastic mean field approach. 
Some experimental evidence for spinodal decomposition has been reported 
by Rivet {\it et al.} \cite{b28}. 

Our model for examining the stability and instability of 
a hot and dilute 
nuclear droplet is based on the diabatic approach to dissipative
large-amplitude collective motion \cite{b29,b30,b31} and, by the use 
of a local energy-density approximation, is similar in spirit to the 
Fermi liquid-drop model of Kolomietz and Shlomo \cite{b24}.
However, instead of using the fluid-dynamical approach with the 
boundary of zero total pressure on the fixed surface as in 
\cite{b24,b43},
we consider a general displacement field, which is defined by an
expansion of the displacement potential in terms of multipoles, and
include Coulomb interactions. This expansion with a suitable boundary
condition for the complete basis set allows the
analytical evaluation of collective mass and stiffness 
tensors within a consistent harmonic approximation. 
The set of eigenvalue equations couple modes with different number of 
nodes in the radial function of the displacement field. 
In contradistinction to \cite{b26},
where dispersion relations of the energy as function of radial wave  
number for different multipoles have been obtained in the adiabatic
limit (no deformations of the local Fermi sphere), we determine the 
orthogonal eigenmodes of the droplet as function of the relaxation time
$\tau$ for the decay of deformations of the local Fermi sphere, {\it i.e.} 
continuously from the adiabatic to the diabatic limit. Furthermore
we study also pure surface modes and compare the instability 
properties for soft and stiff equations of state. 
In this way we extend the exploratory studies \cite{b26,b25,b27,b24}
of spinodal instabilities of finite droplets to a more 
systematic analysis
of stability and instability by examining the eigenmodes of a nuclear
droplet as functions of densities and temperatures, relevant in the
multifragmentation process.

In the following, which forms part I of a series of publications,
we present detailed results on isoscalar bulk and
surface modes with fixed ratio of neutron and proton densities in the
adiabatic limit (very fast equilibration, $\tau\rightarrow 0$).
Since $\tau\propto T^{-2}$, cf.\ \cite{b32}, this corresponds to the 
high-temperature limit. We have studied these adiabatic
iso\-sca\-lar modes in detail, because they are 
related to thermodynamics and to many studies performed in the
past.
In subsequent papers we will present the modifications due to 
dissipation ($0<\tau <\infty$) and extend the study to the inclusion 
of isovector modes, which to our knowledge have been studied so far only 
for infinite nuclear matter \cite{b33,b34,b37*} and for a lattice gas 
\cite{b35}.

The results on adiabatic isoscalar 
bulk instabilities (preliminary calculations have been 
presented earlier \cite{b36}) are summerized as follows.
\begin{itemize}
\item 
As compared to infinite nuclear matter the spinodal region for 
compressional (bulk)
instabilities shrinks to smaller densities and temperatures with 
$T_\mathrm{crit}=6$ MeV for a soft equation of state (EOS). 
This effect is due to an increase of stability,  
which results from the Weizs\"ac\-ker term $\propto (\vgrad\varrho)^2$ 
in the energy density and the finite wave lengths of density  
fluctuations in the droplet. The observed fragmentation temperatures
of about 5 MeV (cf.\ \cite{b17}) are consistent with spinodal 
decomposition after expansion. Typical values for the growth rates 
are $\gamma\approx 5$ MeV corresponding to growth times  
$\hbar/\gamma\approx 40$ fm/c.
\item
Effects from Coulomb interactions on the bulk instabilities are 
negligible, although, as shown in \cite{b37}, they are important at 
later stages of the multifragmentation process, 
where statistical models apply. 
\item
As compared to the soft EOS, 
a stiff EOS yields a larger spinodal region with $T_\mathrm{crit}=8$ 
MeV. Typical values for the growth rates of instabilities are larger 
by almost a factor 2.
\item
With decreasing density and temperature the modes with the lowest
multipolarities and no radial node become unstable first.
\item
At densities below $0.3\varrho_0$ (with $\varrho_0=0.16$ fm$^{-3}$)
the instability growth rates for different multipolarities
$(l=2,3,4,5)$ and number of nodes $(n=0,1,2,3)$ are practically equal.
This property can yield a power-law behavior 
$A^{-\sigma}$ with $\sigma\approx 2.0$ of the fragment-mass distribution
in agreement with experimental observations, cf.~\cite{b38}
and is not related to the critical point. Additional components from
evaporation at lower excitation energies and non-equilibrium 
coalescence at higher excitation energies are considered responsible
for the observed larger values of $\sigma$.
\end{itemize}
For finite nuclear droplets surface modes are important in addition
to the compressional modes. Indeed, pure surface modes show some
interesting features.
\begin{itemize}
\item 
The instability region of pure surface modes extends to larger densities 
up to about the spinodal line of infinite nuclear matter and to large 
temperatures.
\item
In general the growth times are smaller by half an order of magnitude
as compared to the typical values for bulk instabilities.
\item
In the stable region surface modes are slow, such that deformations 
initiated 
in the excitation process will persist during expansion and clustering.
\item
The surface instability is dominated by quadrupole deformation.  
Fission at high excitation 
energies is expected to take this path through a low density stage, 
where the fission barrier vanishes. This is a novel mechanism of fission
and may be even faster than the standard path across the barrier 
at equilibrium density.
\end{itemize}

In the following section \ref{collmodel} we introduce the collective 
model, which serves as the basis for our studies of stability and 
instability of nuclear droplets. 
In section \ref{AIM} we describe the application to
adiabatic isoscalar modes of compressional and pure surface 
deformations.
Results on stability and instability of expanding nuclear droplets are
presented in section \ref{results}. Relations to multifragmentation 
are discussed 
in section \ref{instmul}. Details of derivations and evaluations
are given in the Appendix. 

\section{The collective model}\label{collmodel}

We apply the diabatic approach to dissipative collective motion 
\cite{b29,b30,b31}. In contradistinction to adiabatic models, dynamical
distortions of the Fermi sphere are included and dissipation due to 
two-body collisions give rise to a term in the collective equation of 
motion, which is non-local in time (non-markovian friction).

\subsection{Reminder of the diabatic approach}\label{reminder}

If we introduce a set $\vec{g}\equiv\{g_\lambda \}$ of collective 
variables $g_\lambda(t)$ in the irrotational velocity field
\begin{equation}\label{velfield}
 \vec{v}(\vec{r},t) 
 = \vgrad W(\vec{r},\vec{g},\dot{\vec{g}})
 = \sum_\lambda \dot{g}_\lambda\, \vgrad w_\lambda(\vec{r},\vec{g})
\end{equation}
of collective motion, we can rigorously define (stationary) diabatic 
single-particle states $|\widetilde{\phi}_\alpha(\vec{g})\rangle$ by
\begin{equation}\label{filam}
 \frac{\partial}{\partial g_\lambda}
|\widetilde{\phi}_\alpha(\vec{g})\rangle =
 -\frac{1}{2}\{ (\vgrad w_\lambda)\cdot \vgrad + \vgrad\cdot
 (\vgrad w_\lambda) \}|\widetilde{\phi}_\alpha(\vec{g})\rangle
\end{equation}
{\it i.e.} by an infinitesimal unitary transformation, which deforms 
the wave function in accordance with the velocity field 
(cf. Appendix \ref{ap0}). We search for the solution
$|\Psi(t)\rangle$ of the many-body Schr\"odinger equation by expanding
\begin{equation}\label{Psit} 
 |\Psi(t)\rangle = 
\sum_\nu c_\nu (t) \,|\Psi_\nu (\vec{g},\dot{\vec{g}},t)\rangle
\end{equation}
in terms of Slater determinants $|\Psi_\nu\rangle$, 
which are built from the boosted diabatic single-particle states
\begin{equation}\label{dsinpart} 
 |\widetilde{\psi}_\alpha\rangle \equiv \exp\left\{\frac{i}{\hbar}
 \left[  m_N W(\vec{g},\dot{\vec{g}})-\int_{t_0}^t \rmd t'\, 
 \varepsilon_\alpha (t')
 \right]\right\} |\widetilde{\phi}_\alpha (\vec{g})\rangle
\end{equation}
moving with the velocity field $\vec{v}$. Here, $\varepsilon_\alpha$ and 
$m_N$ denote the diabatic single-particle energies and the nucleon mass,
respectively.

From the variational principle
\begin{equation}\label{varprinc} 
 \delta_{c_\nu,g_\lambda} \int_{t_0}^{t_1} \rmd t \,\langle \Psi(t)
 |i\hbar \frac{\partial}{\partial t} - H|\Psi(t)\rangle = 0
\end{equation}
with the many-body Hamiltonian $H$, one derives equations of motion 
for the expansion coefficients $c_\nu (t)$ and collective variables 
$g_\lambda (t)$. The coupled equations for the expansion coefficients 
describe the mixing of
states (essentially due to residual interactions), and hence the
equilibration process within the intrinsic degrees of freedom.
We approximate this equilibration by a relaxation equation for 
the diabatic
single-particle occupation probabilities
\begin{equation}\label{dnalpha}  
 \frac{\rmd n_\alpha(t)}{\rmd t} = -\frac{1}{\tau(t)} \{n_\alpha(t) -
 \overline{n}_\alpha(\vec{g},\mu,T)\}
\end{equation}
where $\tau$ and $\overline{n}_\alpha$, respectively, 
denote the relaxation time
and the equilibrium values of the occupation probabilities. The chemical
potential $\mu(t)$ and the temperature $T(t)$ in the Fermi distribution for
$\overline{n}_\alpha$ are determined by the conservation of mass and excitation
energy in the relaxation process. The relaxation time $\tau$ has been
estimated for normal nuclear matter by several groups, cf.\ \cite{b39}, and
given as 
\begin{equation}\label{reltime} 
 \frac{\tau}{\hbar} = \frac{\eta}{\varepsilon^*} 
\end{equation}
with values for $\eta$ between 0.15 and 0.30 depending on the magnitude of
the nucleon-nucleon cross-section in medium.

The collective equation of motion, as obtained from the variation 
(\ref{varprinc}) with respect
to $q_\lambda$, is given by \cite{b39*}
\begin{equation}\label{colleq} 
 \frac{\rmd}{\rmd t} \sum_{\lambda\lambda'}\, B_{\lambda\lambda'}\,
 \dot{g}_{\lambda'} - \frac{1}{2} \sum_{\lambda'\lambda''}\,
 \frac{\rmd B_{\lambda'\lambda''}}{\rmd g_\lambda} \dot{g}_{\lambda'}\dot{g}_{\lambda''}
 = F_\lambda \;,
\end{equation}
where $B_{\lambda\lambda'}$ denotes the collective mass tensor 
\begin{equation}\label{masstens} 
 B_{\lambda\lambda'} = m_N \int \rmd^3 r\, \widetilde{\varrho}(\vec{r}) 
 (\vgrad w_{\lambda})\cdot (\vgrad w_{\lambda'})
\end{equation}
for the velocity field (\ref{velfield}) and the density 
\begin{equation}\label{tilden} 
\widetilde{\varrho}(\vec{r})
= \sum_\alpha n_\alpha|\widetilde{\phi}_\alpha (\vec{r})|^2\;.
\end{equation}
The force on the r.h.s.\ of (\ref{colleq}) is determined by
\begin{equation}\label{force}
 F_\lambda = -\sum_\alpha \frac{\partial \varepsilon_\alpha (\vec{g})}
 {\partial g_\lambda}\, n_\alpha (t) 
\end{equation}
in the diabatic representation. Note that the diabatic single-particle
energies are smooth functions of $\vec{g}$ and that the force is defined for
fixed $n_\alpha$, {\it i.e.} for constant (single-particle) entropy.

Inserting in (\ref{force}) the formal integral of (\ref{dnalpha})
\begin{equation}\label{nalphat}
 n_\alpha(t)= \overline{n}_\alpha(t) - \int_{t_0}^t \rmd t'
 \,\frac{\rmd \overline{n}_\alpha}{\rmd t'}\,  e^{-\int_{t'}^t\, \rmd t''/\tau(t'')}
\end{equation}
with the initial value  $n_\alpha(t_0)=\overline{n}_\alpha (t_0)$, we obtain
\begin{equation}\label{flamb}
 F_\lambda (\vec{g},t)=\overline{F}_\lambda (\vec{g},t)+F'_\lambda (\vec{g},t)\;,
\end{equation}
\begin{equation}\label{flambbar}
 \overline{F}_\lambda (\vec{g},t) = -\sum_\alpha 
 \frac{\partial \varepsilon_\alpha (\vec{g})}
 {\partial g_\lambda}\,\overline{n}_\alpha\;,
\end{equation}
\begin{equation}\label{flambprim}
 F'_\lambda (\vec{g},t) =  \sum_\alpha 
 \frac{\partial \varepsilon_\alpha (\vec{g})}
 {\partial g_\lambda} \int_{t_0}^t \rmd t'\,
\frac{\rmd \overline{n}_\alpha}{\rmd t'} 
 \, e^{-\int_{t'}^t\, \rmd t''/\tau(t'')} .
\end{equation}
The equilibrium force is determined by the equilibrium distribution
$\overline{n}_\alpha (t)$ for the time-dependent temperature and 
chemical potential. 
For temperatures ($T\approx 1$ MeV) large compared to the 
residual single-particle coupling, {\em i.e.} difference
between the total mean field potential (adiabatic potential) and the 
diabatic potential, $\overline{F}_\lambda$ is identical to the adiabatic 
force, because the sum of neighboring energies 
is the same in both cases (invariance of trace). 
The elastoplastic force $F'_\lambda$ 
describes giant elastic vibrations with frequency 
$\omega$ for $\tau \gg \omega^{-1}$ and Markov dissipation 
(ordinary friction) for $\tau \ll \omega^{-1}$. 
Whereas large-amplitude collective motion has been considered in 
\cite{b29,b30,b31}, we specify the following discussion to 
small amplitudes. 

\subsection{Small-amplitude modes}\label{smallamp}

For small amplitudes around some equilibrium point $\vec{g}_0$=0 
we keep in (\ref{colleq}) only terms linear in 
$g_\lambda$, and hence we have 
\begin{equation}\label{eom}
\sum_{\lambda'}\, B_{\lambda\lambda'} \ddot{g}_{\lambda'} + 
 C'_{\lambda\lambda'}
 \int_{t_0}^t \rmd t'\, e^{-(t-t')/\tau}\,\dot{g}_{\lambda'}(t') +
 \overline{C}_{\lambda\lambda'}\, g_{\lambda'} = 0
\end{equation}
with $B_{\lambda\lambda'}=B_{\lambda\lambda'}(\vec{g}_0)$, 
$\tau=\tau (t_0)$ and the stiffness tensors 
\begin{equation}\label{barC}
 \overline{C}_{\lambda\lambda'}= \sum_\alpha \left( \frac{\partial^2 
 \varepsilon_\alpha}{\partial g_{\lambda}\partial g_{\lambda'}}\,
 \overline{n}_\alpha
 + \frac{\partial \varepsilon_\alpha}{\partial g_{\lambda}} 
  \frac{\partial \overline{n}_\alpha}{\partial g_{\lambda'}}
 \right)_{\vec{g}=0} \;,
\end{equation}
\begin{equation}\label{C'}
 C'_{\lambda\lambda'} = - \sum_\alpha \left( 
 \frac{\partial \varepsilon_\alpha} {\partial g_{\lambda}} \, 
 \frac{\partial \overline{n}_\alpha}{\partial g_{\lambda'}}
 \right)_{\vec{g}=0} \;.
\end{equation}
Inserting $g_{\lambda} \propto \exp(-i\omega t)$, we obtain from 
(\ref{eom}) the eigenvalue equation 
\begin{equation}\label{eigenveq}
 G^{-1}_{\lambda\lambda'}(\omega) = -B_{\lambda\lambda'}\,\omega^2 +
 C'_{\lambda\lambda'}\frac{\omega}{\omega+ i/\tau} +
 \overline{C}_{\lambda\lambda'} = 0 
\end{equation}
for $t-t_0\gg \tau$, 
where $G_{\lambda\lambda'}(\omega)$ denotes the Fourier transform 
of the Green function for (\ref{eom}). In general, {\it i.e.} 
for finite $\tau$-values, there are three complex solutions to 
this equation. For the two elastic (isentropic) limits, 
$\tau=0$ (adiabatic limit) and $\tau\rightarrow\infty$ 
(diabatic limit) only two, either real ($\omega^2 > 0$) or 
imaginary ($\omega^2 < 0$) frequencies survive. In the diabatic limit
the stiffness tensor is given by 
\begin{equation}\label{stifftens}
 C_{\lambda\lambda'} = C'_{\lambda\lambda'} + 
 \overline{C}_{\lambda\lambda'}
 = \sum_\alpha \left( \frac{\partial^2 \varepsilon_\alpha}
 {\partial g_{\lambda} \partial g_{\lambda'}}\right)_{\vec{g}=0}
 \overline{n}_\alpha \;,
\end{equation}
whereas it is $\overline{C}_{\lambda\lambda'}$ in the adiabatic limit. 
As mentioned below (\ref{flambprim}) the adiabatic stiffness 
tensor can be expressed for temperatures $T\gtrsim 1$ MeV by
\begin{equation}\label{stifftensad}
 \overline{C}_{\lambda\lambda'} = \sum_{\overline{\alpha}} \left( 
 \frac{\partial^2 \varepsilon_{\overline{\alpha}}}
 {\partial g_{\lambda} \partial g_{\lambda'}} \right)_{\vec{g}=0}
 n_{\overline{\alpha}} \;,
\end{equation}
where $\overline{\alpha}$ denote the adiabatic single-particle states.

\subsection{Local-density approximation}\label{locdensap}

In applying the formulation to the modes of a spherical nuclear 
droplet of homogeneous density $\varrho$ and temperature $T$ 
inside the sharp surface, we perform a local density
approximation. The mass tensor (\ref{masstens}) is already given in an
appropriate form. For the stiffness tensors (\ref{stifftens}) and 
(\ref{stifftensad}) we make use of the relation
\begin{eqnarray}\label{relation}
 \sum_\alpha \frac{\partial \varepsilon_\alpha}{\partial g_{\lambda}}\,
 n_{\alpha} & = & \sum_\alpha \frac{\partial t_{\alpha\alpha}}
{\partial g_{\lambda}}\,n_{\alpha} + 
 \frac{\partial}{\partial g_{\lambda}}\,
 \frac{1}{2} \sum_{\alpha\beta} v^\mathrm{as}_{\alpha\beta\alpha\beta}\, 
 n_{\beta} n_{\alpha} \nonumber \\ & = & \frac{\partial} 
 {\partial g_\lambda} \left( E_{\mathrm{kin}}(\vec{g}) 
 + E_{\mathrm{int}}(\vec{g}) \right) \;,
\end{eqnarray}
where $t_{\alpha\alpha}$ and $v^\mathrm{as}_{\alpha\beta\alpha\beta}$ 
denote the kinetic-energy part of the single-particle energy 
and the antisymmetrized two-body interaction matrix element, 
respectively. The total intrinsic kinetic energy 
$E_{\mathrm{kin}}(\vec{g})$ and the interaction energy 
$E_{\mathrm{int}}(\vec{g})$ can be  expressed by Skyrme functionals 
$\epsilon[\widetilde{\varrho}]$ for the energy density, where 
$\widetilde{\varrho}$ denotes the deformed density due to collective
motion. Furthermore, we introduce a phenomenological surface energy 
tension $\epsilon_S(\widetilde{\varrho})$. 
Then the stiffness tensors are defined by 
\begin{equation}\label{stifftSK}
 C_{\lambda\lambda'} = \left\{ \frac{\partial^2}
 {\partial g_{\lambda}\partial g_{\lambda'}}\left[ \int \rmd^3r\, 
 \epsilon[\widetilde{\varrho}] + \int \rmd f\,
 \epsilon_S(\widetilde{\varrho}) \right]\right \}_{\vec{g}=0} ,
\end{equation}
\begin{equation}\label{stifftSKbar}
 \overline{C}_{\lambda\lambda'} = \left\{ \frac{\partial^2}
 {\partial g_{\lambda}\partial g_{\lambda'}}\left[ \int \rmd^3r\, 
 \overline{\epsilon}[\widetilde{\varrho}] + \int \rmd f\,
 \epsilon_S(\widetilde{\varrho}) \right]\right \}_{\vec{g}=0}  
\end{equation}
in the local density approximation, where $\epsilon$ differs from the
adiabatic value $\overline{\epsilon}$ by the additional deformation 
of the local Fermi sphere in the diabatic limit. 
Note that the derivatives in (\ref{stifftSK}) and (\ref{stifftSKbar}) 
have to be taken at constant entropy, {\em i.e.} for
fixed occupation probabilities in accordance with 
(\ref{stifftens}) and (\ref{stifftensad}). 

\subsection{Energy densities and surface tension}\label{Edast}

We write the Skyrme energy-density functional \cite{b40} as a sum 
\begin{equation}\label{gf16}
 \epsilon[\widetilde{\varrho}] = \epsilon_{\tau} + \epsilon_V + 
 \epsilon_W + \epsilon_C
\end{equation}
of terms related to the intrinsic kinetic energy ($\epsilon_{\tau}$),
including the momentum-dependent part of the interaction,
the nuclear interaction of homogeneous systems ($\epsilon_V$), 
the contribution from 
inhomogeneity (Weizs\"acker term $\epsilon_W$) and the Coulomb 
interaction ($\epsilon_C$). Explicitly these terms read
\begin{equation}\label{gf17}
\epsilon_{\tau} =  \epsilon_{\tau}^{(n)} + \epsilon_{\tau}^{(p)} \;,
\end{equation}
\begin{equation}\label{gf17a}
\epsilon_{\tau}^{(i)} = g\, \int \frac{\mathrm{d}^3k}{(2\pi)^3}\,
\frac{\hbar^2 k^2}{2m_i^*}\,\widetilde{f}_i(k)\;,
\end{equation}
\begin{eqnarray}\label{gf18}
\epsilon_V & = & \frac{1}{2} t_0 \left[ 
 (1+ \frac{1}{2}x_0) \widetilde{\varrho}^2 
 - (x_0+\frac{1}{2}) (\widetilde{\varrho}_n^2 + 
 \widetilde{\varrho}_p^2) \right]\\
 &+& \frac{1}{12} t_3 \widetilde{\varrho}^{\alpha} \left[ 
 (1+\frac{1}{2}x_3)\widetilde{\varrho}^2
 -(x_3+\frac{1}{2})(\widetilde{\varrho}_n^2+\widetilde{\varrho}_p^2) 
 \right]  \; ,\nonumber
\end{eqnarray}
\begin{eqnarray}\label{gf19}
 \epsilon_W & = & \frac{1}{16} (3t_1-t_2) 
 (\nabla \widetilde{\varrho})^2 \\ 
 & + &  \frac{1}{32} (3t_1+t_2) \left[ 
 (\nabla \widetilde{\varrho}_n)^2 + 
 (\nabla \widetilde{\varrho}_p)^2 \right] \; , \nonumber
\end{eqnarray}
\begin{equation}\label{gf19a}
 \epsilon_C = \frac{1}{2}\, \int \mathrm{d}^3 r' \,
 \frac{e_0^2\,\widetilde{\varrho}_p(\vec{r})
 \widetilde{\varrho}_p(\vec{r}')} {|\vec{r}-\vec{r}'|}  \; ,
\end{equation}
where $g=2$ is the spin degeneracy factor, $\widetilde{f}_i(k)$ 
denote the local momentum distributions (normalized to density) 
for neutrons ($i=n$) and protons ($i=p$) inside the droplet and 
$m_i^*$ the corresponding effective masses given by 
\begin{equation}\label{gf20}
\frac{2\hbar^2}{m_i^*} = \frac{2\hbar^2}{m_N} +[(t_1+t_2)\widetilde{\varrho}+\frac{1}{2}
(t_2-t_1)\widetilde{\varrho}_i] \; .
\end{equation}
Note that the $\vec{r}$-dependence in eqs.\ (\ref{gf17}-\ref{gf19})
and (\ref{gf20}) is not explicitly indicated.

Two Skyrme forces SkM$^*$ and SIII are considered, which correspond 
to soft and stiff equations of state, respectively. 
The parameters are given in Table \ref{Skyf}.
 \begin{table}[tbh]
 \vspace{-2ex}
 \caption{Parameters of SIII and SkM$^*$ Skyrme forces}
 \label{Skyf}
 \begin{center}
 \begin{tabular}{|l||c|c|}\hline\
 & SIII & SkM$^*$ \\ \hline\hline
 $t_0$ (MeV fm$^3$) & -1128.75 & -2645 \\ \hline
 $t_1$ (MeV fm$^5$) & 395 & 410 \\ \hline
 $t_2$ (MeV fm$^5$) & -95 & -135 \\ \hline
 $t_3$ (MeV fm$^{3(1+\alpha)}$) & 14000 & 15595 \\ \hline
 $x_0$ & 0.45 & 0.09 \\ \hline
 $x_3$ & 1 & 0 \\ \hline
 $\alpha$ & 1 & 1/6 \\ \hline
 \end{tabular}
 \end{center}
 \vspace{-2ex}
 \end{table}
 
The surface energy is  
directly determined from the surface tension 
(cf.\ Appendix \ref{bb} and \cite{b41,b42})
\begin{equation}\label{gf21}
\epsilon_S(\widetilde{\varrho}) = \frac{a_s}{4\pi r_0^2} 
\left(\frac{\widetilde{\varrho}}{\varrho_\mathrm{eq}}\right)^2 
\,(1-a_{\delta}\xi^2)\, (1+\beta T^2) \:,
\end{equation}
where $a_s = 17$ MeV, $r_0=1.2$ fm, 
$a_{\delta}=(3t_1+t_2)/(9t_1-5t_2)$, $\xi=(N-Z)/A$ and
$\beta=0.006$ MeV$^{-2}$ or 0.008 MeV$^{-2}$ for the soft
and stiff EOS, respectively.
Here, $\varrho_\mathrm{eq}$ denotes the equilibrium nuclear density
of the homogeneous nuclear droplet. For not too light nuclei
the value is $\varrho_\mathrm{eq}\approx0.85\,\varrho_0$ with
$\varrho_0=0.16$ fm$^{-3}$. 

\section{Adiabatic isoscalar modes}\label{AIM}

In the adiabatic limit ($\tau=0$) the eigenvalue equation (\ref{eigenveq}) 
reduces to 
\begin{equation}\label{gf13}
\overline{C}_{\lambda \lambda'}-\omega^2\,B_{\lambda \lambda'}=0 
\end{equation}
with the adiabatic stiffness tensor given by 
(\ref{stifftSKbar}) in the local-density approximation. 
For ${\omega}^2>0$  the corresponding mode is stable 
(${q}_{\lambda} \propto \sin(\omega t)$). For ${\omega}^2= -\gamma^2<0$
the mode is unstable, {\it i.e.} a small fluctuation leads to 
an exponential growth of the amplitude 
(${q}_{\lambda} \propto \exp(\gamma t)$).

Since we are restricting now our study to isoscalar modes, the neutron
and proton densities are given by 
$\widetilde{\varrho}_n=\widetilde{\varrho}\, N/A$ and
$\widetilde{\varrho}_p=\widetilde{\varrho}\, Z/A$ with $A=N+Z$.

In the following we treat the compressional modes and the pure surface modes 
separately.

\subsection{Compressional modes}\label{bulk}

Collective variables are introduced by the coefficients of the
expansion of an irrotational displacement field in 
terms of a complete set of
functions. Analytical expressions are derived for the mass and 
stiffness tensors.  

\subsubsection{Collective variables}\label{general}

We define a set of real functions for $r\leq R$ 
\begin{equation}\label{gf1}
\PsilrO = \Nln \, \jlkr\, \cYlmnO 
\end{equation}
with $\lambda \equiv \{nlm\}$.
Here, $\jlkr$ denote the spherical Bessel functions and $\cYlmnO$ 
real spherical harmonics defined from the complex ones 
$Y_{l}^{\mu}(\Omega)$ by
\begin{eqnarray}\label{gf2}
{\cal Y}_{l}^{m} = \left\{ \begin{array}{l}
	\sqrt{\frac{m}{2|m|}}
	[Y_{l}^{|m|} + \frac{m}{|m|}
	{Y_{l}^{|m|}}^*],
	\hspace{1ex}  \mbox{for}\hspace{1ex}m\neq 0\\
	Y_{l}^{0},  \hspace{22.7ex} \mbox{for}\hspace{1ex}m=0 
	\end{array} \right. 
\end{eqnarray}
with the boundary condition 
\begin{equation}\label{gf3}
\chi_\lambda(r=R,\Omega)= j_l(\kappa_{nl}R)=0 
\end{equation}
($n=0,1,2,\ldots$) and the normalization constant
\begin{equation}\label{gf4}
\Nln=\sqrt{\frac{2}{R^3}}\,\frac{1}{j'_l(\kappa_{nl}R)} 
= \sqrt{\frac{2}{R^3}}\,\frac{1}{j_{l-1}(\kappa_{nl}R)} \,.
\end{equation}
The basis (\ref{gf1}) forms a complete set of orthonormal functions 
within $r\leq R$ satisfying 
\begin{equation}\label{gf5}
\Delta \chi_{\lambda} = -\kappa_{nl}^2 \,\chi_{\lambda}\;, 
\end{equation}
\begin{equation}\label{gf6}
\int \mathrm{d}^3 r\, \chi_{\lambda}(\vec{r})\,\chi_{\lambda'}(\vec{r}) =
\delta_{\lambda \,\lambda' }\; .
\end{equation}
\begin{figure}[tb]
\begin{center}
 \resizebox{0.48\textwidth}{!}{ \includegraphics{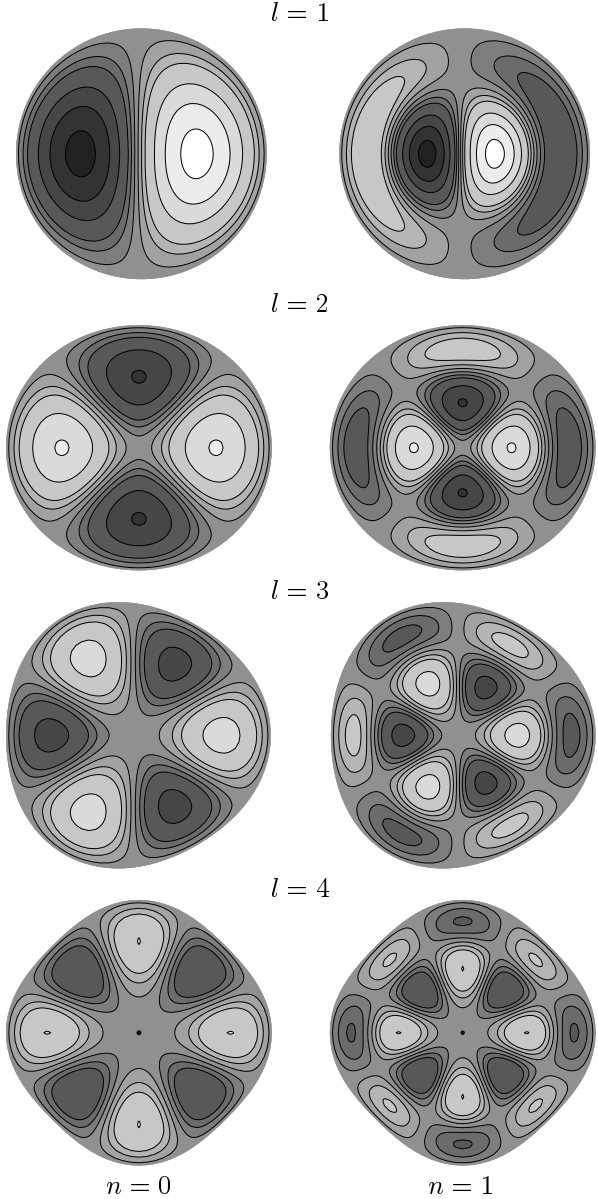}  } 
\end{center}
\caption{Density distributions (cut in the $(x,y)$--plane)
of a droplet for the 
lowest collective modes with $n=0$ and 1 (left and right column,
respectively) and $l=1$ to 4 (from top to bottom).
Darker shades correspond to larger densities, lighter ones to 
smaller densities.
The grey color in the center and at the border corresponds to 
the density of the undistorted sphere.}
\vspace*{-2mm}
\label{modshap}       
\end{figure}

Deformations of the spherical droplet with homogenous density $\varrho$
and radius $R$ are conveniently described by a displacement field 
$\vec{s}(\vec{r},t)$.
We introduce collective variables by the expansion coefficients 
$q_{\lambda}(t)$ of the displacement potential 
\begin{equation}\label{gf7}
w(\vec{r},t) = \sum_{\lambda}\,
q_{\lambda}(t)\,\chi_{\lambda}(\vec{r}) 
\end{equation}
and the corresponding irrotational displacement field 
\begin{equation}\label{gf8}
 \vec{s}(\vec{r},t) = \vgrad \, w(\vec{r},t) = 
 \sum_{\lambda}\,q_{\lambda}(t)\,\vgrad\chi_{\lambda}(\vec{r}) \; .
\end{equation}
Note that $q_{\lambda}$ has the dimension of (length)$^{7/2}$.
The time derivative
\begin{equation}\label{gf9}
\dot{\vec{s}}(\vec{r},t) =  \sum_{\lambda}\,
\dot{q}_{\lambda}(t)\,\vgrad \chi_{\lambda}(\vec{r}) 
\end{equation}
determines the velocity field $\vec{v}$ at the position 
$\vec{r}+\vec{s}(\vec{r},t)$.
This shift $\vec{s}(\vec{r},t)$ makes our expansion different from 
the one for the irrotational velocity field (\ref{velfield}), 
which is commonly used (cf.\ \cite{b43}). 
However, for $\vec{s}=0$ we have $\dot{\vec{s}}=\vec{v}(\vec{r})$, and hence
all terms in (\ref{dsinpart}), (\ref{varprinc}), (\ref{colleq}) 
and (\ref{masstens}), which are related to the collective kinetic energy, 
remain unchanged. Since for small amplitudes 
$B_{\lambda \lambda'}=B_{\lambda \lambda'}(\vec{q}_0=0)$, the kinetic 
energy part in (\ref{eom}) is not affected by the difference in the expansion.
Of course, the stiffness coefficients are different due to the difference 
in the displacement fields. We have chosen the irrotational displacement field
(\ref{gf8}), because it conserves the center of mass exactly for all $l$-values 
including $l=1$ (cf.\ Appendix \ref{ap2}), and furthermore allows to calculate
mass and stiffness tensors analytically.

The density $\widetilde{\varrho}(\vec{r},t)$
varies according to the continuity equation
\begin{equation}\label{gf10}
\frac{\partial \widetilde{\varrho}(\vec{r},t)}{\partial t} +
\mbox{div} [\widetilde{\varrho}(\vec{r},t)\,\vec{v}(\vec{r},t)] = 0\; ,
\end{equation}
which assures mass conservation. 
Figure~\ref{modshap} illustrates shapes and density profiles 
(calculated from eq.~(\ref{a0}) of Appendix \ref{ap3})
for the lowest collective modes with $n=0,1$ and $l=1,2,3,4$. 
Since we want to describe the fragmentation into at least two large
fragments, we restrict the expansion  (\ref{gf7}) to $l\geq 2$.
Instabilities with $l=0$ and  $l=1$ are expected to lead essentially
to a single heavy fragment with additional light particles 
($n, p, d, \alpha$\ldots), and hence can not be easily distinguished
from evaporation events. 

As discussed in Appendix \ref{ap3}, the homogeneous sphere $(\vec{q}=0)$
is an equilibrium point in the space of 
the collective variables $q_{\lambda}$ for $l>0$, and hence (\ref{eom})  
with $g_\lambda\rightarrow q_\lambda$ describes its harmonic modes. 

\subsubsection{Mass tensor}\label{explicit}

The mass (inertial) tensor $B_{\lambda \lambda'}$ is calculated
from the collective kinetic energy tensor. 
Every mass element $m_N\varrho\,\mathrm{d}^3 r$ of the unperturbed
droplet contributes 
$\frac{1}{2} m_N\varrho\,\mathrm{d}^3r[\dot{\vec{s}}(\vec{r},t)]^2$ 
to the kinetic energy (cf.\ (\ref{aa7p})),
where $m_N$ denotes the nucleon mass and $\varrho$ the constant value
of the density inside the unperturbed sphere of radius $R$. 
Inserting the expansion 
(\ref{gf9}) for $\dot{\vec{s}}(\vec{r},t)$ we obtain for the
inertial tensor
\begin{eqnarray}\label{gf15}
B_{\lambda \lambda'} &=& m_N \, \varrho \int_{r\leq R} \mathrm{d}^3 r \,
(\vgrad \chi_{\lambda})\cdot (\vgrad \chi_{\lambda'}) \nonumber \\
& = & \delta_{\lambda \lambda'} \, m_N \, \varrho \, \kappa_{nl}^2 
\end{eqnarray}
according to the definition (\ref{masstens}), where the final expression 
is diagonal and 
results from integrating by parts and using (\ref{gf3}) and (\ref{gf5}).

\subsubsection{Stiffness tensor}\label{stiffness}

According to the different energy contributions (intrinsic 
kinetic energy, local and nonlocal nuclear interaction, 
Coulomb and surface energies, cf.\ sect.\ \ref{Edast}), 
the adiabatic stiffness tensor
\begin{equation}\label{gf22}
\Cllbar = \Clltbar + \CllV + \CllW+ \CllC + \CllS   
\end{equation}
is the sum of five terms.
The derivation  of explicit formulae for the individual terms is 
straightforward. We refer to Appendix \ref{aa} for details. 

The intrinsic kinetic energy term $\Clltbar$ is calculated in
the adiabatic limit, which is defined by leaving the occupation 
of the lowest single-particle levels unchanged. 
The final expression is
\begin{eqnarray}\label{gf23}
\Clltbar &=& \sum_{i=n,p} \, 2\epsilon_{\tau}^{(i)}(\varrho,T) 
\left \{ \left (
\frac{5}{9}+ \frac{5}{3}\mu^{(i)} \right )\kappa_{nl}^4 \,\dll \right.
\nonumber \\ 
&-& \left.\left (\frac{4}{3}+2\mu^{(i)} \right )\frac{\kappa_{nl}\kappa_{n'l}}
{R^2}\,\delta_{ll'}\delta_{mm'} \right \} 
\end{eqnarray}
with
\begin{equation}\label{gf25}
\mu^{(i)} = \varrho\, m_i^*(\varrho)
\frac{\mathrm{d}}{\mathrm{d}\varrho}
\frac{1}{m_i^*(\varrho)}  \;. 
\end{equation}
The kinetic energy density $\epsilon_{\tau}^{(i)}(\varrho,T)$ is defined by
(\ref{gf17a}) and may be approximated by 
\begin{equation}\label{46a}
\epsilon_{\tau}^{(i)}(\varrho,T) \approx  \epsilon_{\tau}^{(i)}(\varrho,0)
\,(1+\alpha T^2) 
\end{equation}
with 
\begin{equation}\label{46b}
\epsilon_{\tau}^{(i)}(\varrho,0) = \frac{3}{20}(3\pi^2)^{2/3}\,
\frac{\hbar^2\,\varrho_i^{5/3}}{m_i^*(\varrho)} \; ,
\end{equation} 
\begin{equation}\label{46c}
\alpha = \frac{5}{3}\left( \frac{1}{9\pi}\right)^{2/3} 
\left( \frac{m_i^*(\varrho)}{\hbar^2} \right)^2 \varrho_i^{-4/3}
\end{equation}
for not too large temperatures $(T\ll \epsilon_{F}^{(i)})$, cf.\ \cite{b40*}.
The results, reported in sect.\ \ref{results} have been obtained 
by using the exact expression (\ref{gf17a}).
Since $(\kappa_{nl}R)^2 \gg 1$, the first (diagonal) term is the largest
contribution in (\ref{gf23}).
 
The contribution from the nuclear-interaction density $\epsilon_V$ 
is obtained as 
\begin{eqnarray}\label{gf26}
\CllV &=& {\varrho}^2 \frac{\mathrm{d}^2 \epsilon_V}
{\mathrm{d}\varrho^2}\, \kappa_{nl}^4\,\dll \\
&-& \left \{ \varrho \frac{\mathrm{d} \epsilon_V}{\mathrm{d}\varrho}
-\epsilon_V(\varrho) \right \}
\frac{4\kappa_{nl}\kappa_{n'l}}{R^2}\,\delta_{ll'}\delta_{mm'} 
\, .\nonumber
\end{eqnarray}
The contributions from the Weizs\"acker and Coulomb terms 
of the energy density are diagonal and read
\begin{equation}\label{gf27}
\CllW = \frac{\mathrm{d}^2\epsilon_W}{\mathrm{d}(\nabla \varrho)^2}\, 
{\varrho}^2\,\kappa_{nl}^6 \,\dll \; ,
\end{equation}
\begin{equation}\label{gf28}
\CllC = \frac{8\pi}{3}\,e_0^2\left (\frac{Z}{A}\varrho \right)^2
\kappa_{nl}^2 \,\dll \;,
\end{equation}
where $e_0$, $Z$ and $A$ denote the elementary charge and the
charge and mass numbers of the nucleus, respectively.

For the surface energy contribution to the stiffness tensor we find
\begin{equation}\label{gf29}
 \CllS = 4\epsilon_S(\varrho) \left\{ 9 + \frac{1}{2}\,l(l+1)
  \right\} \,
  \frac{\kappa_{nl}\kappa_{n'l}}{R^3} 
\,\delta_{ll'}\delta_{mm'}\;\; .
\end{equation}
The stiffness tensor, as given by expressions (\ref{gf23}),(\ref{gf25}) 
and (\ref{gf26}) to 
(\ref{gf29}), is diagonal in $l$ and $m$ and not depending on $m$.
The only couplings left are those corresponding to different 
$n$ values for the same multipolarity and are due to 
$\Clltbar, \CllV$ and $\CllS$.

\begin{figure}[t]
\begin{center}
\resizebox{0.48\textwidth}{!}{%
  \includegraphics{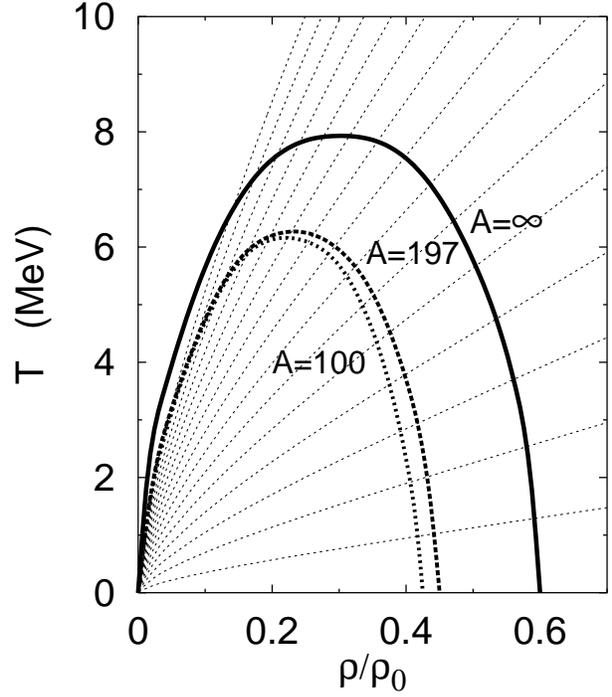} }  \end{center}
\caption{Borders of compressional adiabatic spinodal instabilities 
in the $(\varrho,T)$-plane
for infinite symmetric nuclear matter ($A$=$\infty$, heavy solid line),
gold-like ($A$=197, $Z$=79, $N$=118, dashed line) 
and tin-like ($A$=100, $Z$=$N$=50, dotted line) nuclear droplets
for a soft EOS (SkM$^*$). Here, $\varrho_0=0.16$ fm$^{-3}$
is the normal nuclear matter density. Note that finite nuclei have
central densities around $0.85\varrho_0$. The thin dashed lines 
indicate the expansion trajectories in the $(\varrho,T)$-plane
at constant entropy  
$T\propto \varrho^{2/3}/m^*(\varrho)$ for infinite nuclear matter.}
\label{l2finsiz}       
\end{figure}

\subsubsection{Infinite nuclear matter}\label{infsssc}

The relation to thermodynamic properties of infinite 
nuclear matter is obtained by discarding Coul\-omb interactions and by
taking the limit $R\to\infty$. Then $C^{(C)}$ and $C^{(S)}$ are negligible
and $\kappa_{nl}\rightarrow \kappa$ is no longer restricted by (\ref{gf3})
to discrete values. For any finite $\kappa$ we have 
$\kappa R\rightarrow\infty$, and hence
\begin{equation} \label{clalap}
\Cllbar= (c_1 + c_2\,\kappa^2)\kappa^4\,\delta_{\lambda\lambda'}
\end{equation} 
with 
\begin{equation} \label{c1c2}
c_1=\varrho\left(\frac{\mathrm{d}^2\overline{\epsilon}}{\mathrm{d}\varrho^2}
\right),\quad\quad c_2=\varrho^2\left(\frac{\mathrm{d}^2\epsilon_W}
{\mathrm{d}(\nabla\varrho)^2} \right),
\end{equation}
$\overline{\epsilon}$ denoting the sum of the adiabatic kinetic and 
interaction energies.
The eigenvalues are given by the dispersion relation
\begin{equation} \label{omeg2}
\omega^2= \Cllbar/B_{\lambda\lambda} =
\kappa^2(c_1 + c_2\,\kappa^2)/(m_N\varrho) \;,
\end{equation}
a well-known result in infinite matter (cf. \cite{b44}).
In the adiabatic spinodal regime, {\it i.e.} in the $(\varrho,T)$-plane,
where 
$c_1=\varrho (\mathrm{d}^2\epsilon/\mathrm{d}\varrho^2) <0$, 
unstable modes with $\omega^2<0$ exist according to (\ref{omeg2})
for finite values of $\kappa$ up to a critical value 
$\kappa_\mathrm{crit}=\sqrt{-c_1/c_2}$. The border of the spinodal 
regime is determined by $\kappa_\mathrm{crit}=0$, {\it i.e.} $c_1=0$.
At this point we immediately understand the large reduction of the 
spinodal region for finite systems (cf.\ fig. \ref{l2finsiz}) by the 
finite values for the smallest $\kappa_{nl}$ determined by (\ref{gf3}).

\subsection{Pure surface modes}\label{Psm}

Pure surface modes, without any change of density inside, 
cannot be described by the expansion (\ref{gf8}) for the 
displacement field $\vec{s}$. Instead, one has to use an expansion for 
the velocity field $\vec{v}$ satisfying $\vgrad\cdot\vec{v}=0$, 
such that the continuity equation (\ref{gf10}) yields 
$\partial \widetilde{\varrho}/\partial t=0$ if 
$\widetilde{\varrho}= \varrho = const$ inside, initially. 
This problem is well studied and we essentially quote the
results from \cite{b43}.

\begin{figure}[t]
\begin{center}
\resizebox{0.48\textwidth}{!}{%
  \includegraphics{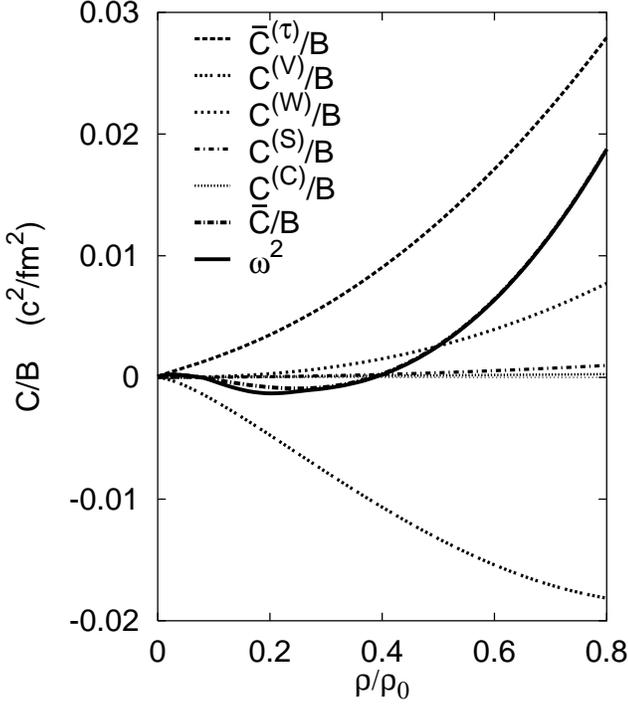} }  \end{center}
\caption{Contributions from the different terms (cf.\ eq.\ (\ref{gf22}))
to the diagonal element of $B^{-1}C$ with $l=2$, $n=0$ for a hot  
($T=4$ MeV) Au-like droplet ($Z,A=79,197)$ described by a  
soft EOS (SkM$^*$). 
Also shown is the lowest eigenvalue $\omega^2$, which differs from 
$\overline{C}/B$ due to couplings with modes $n>0$.}
\label{l2CBa}       
\end{figure}
\begin{figure}[t]
\begin{center}
\resizebox{0.48\textwidth}{!}{%
  \includegraphics{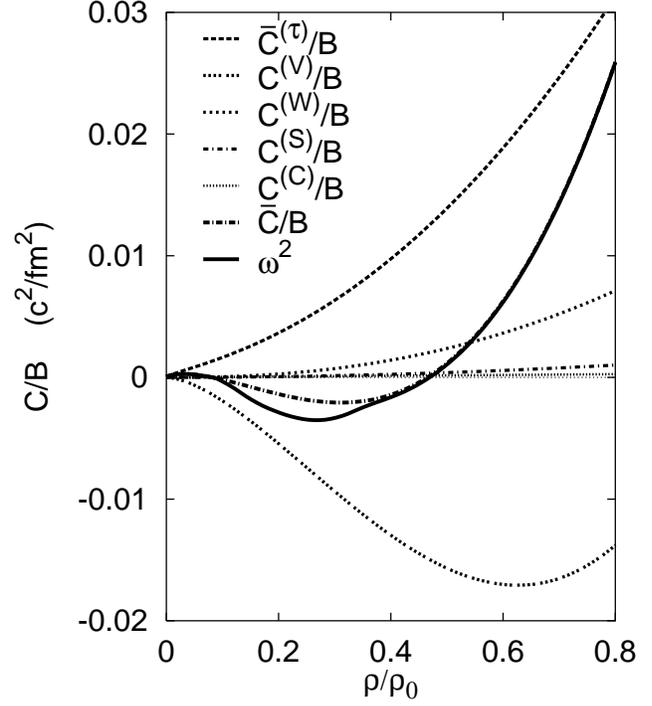} }  \end{center}
\caption{The same as in fig. \ref{l2CBa}, but for a stiff EOS (SIII).}
\label{l2CBahard}       
\end{figure}

\subsubsection{Collective variables}\label{CvBM}

It is convenient to introduce real collective variables $Q_{lm}$, which 
describe the surface $r(\Omega)$ of the deformed droplet by the expansion
\begin{equation} \label{romeg}
 r(\Omega,t) = R\left( 1+\sum_{lm} Q_{lm} {\cal Y}^m_l(\Omega)\right)\;,
\end{equation}
where we use the real spherical harmonics (\ref{gf2}) instead of the 
complex $Y^m_l(\Omega)$. The corresponding velocity field, which 
satisfies $\vgrad\cdot\vec{v}=0$, is defined by $v_r=\dot{r}$ at the 
undeformed surface, and hence by the irrotational field
\begin{equation} \label{irrfield}
 \vec{v}(\vec{r},t) = \sum_{lm} \frac{\dot{Q}_{lm}}{l\, R^{l-2}}\vgrad 
 (r^l{\cal Y}^m_l)\;.
\end{equation}

\subsubsection{Mass tensor}\label{MtBM}

The collective mass tensor is obtained from (\ref{gf15}) by a partial 
integration, using $\Delta (r^l{\cal Y}^m_l)=0$ and the Gauss integral 
formula. This yields the expression 
\begin{equation} \label{mtenBM}
 B_{lm,l'm'}= \delta_{ll'}\delta_{mm'}\, m_N \varrho \frac{R^5}{l} \; , 
\end{equation}
where the $\vec{Q}$-dependence is neglected (harmonic or small-amplitude
approximation).

\subsubsection{Stiffness tensor}\label{StBM}

Since the density remains constant during the surface deformations, 
only the surface and
Coulomb energies contribute to the stiffness tensor, {\it i.e.}
\begin{equation} \label{stenBM}
 C_{lm,l'm'} = C_{lm,l'm'}^{(S)} + C_{lm,l'm'}^{(C)} \;.
\end{equation}
In applying the expressions of \cite{b43} we note that
\begin{equation} \label{alphalm}
 \alpha_{lm} = \frac{1}{\sqrt{2}}(Q_{lm}+iQ_{l-m}) \;,
\end{equation}
and hence $\sum_{m} |\alpha_{lm}|^2 +|\alpha_{l-m}|^2 = 
\sum_{m} Q_{lm}^2 + Q_{l-m}^2$. Thus, 
also the stiffness tensors have to be equal to those of \cite{b43},
\begin{equation} \label{SstenBM}
 C_{lm,l'm'}^{(S)} = \delta_{ll'}\delta_{mm'}\, \frac{(l-1)(l+2)}{4\pi}\,
 a_s A^{2/3} \;,
\end{equation}
\begin{equation} \label{CstenBM}
 C_{lm,l'm'}^{(C)} = - \delta_{ll'}\delta_{mm'}\, \frac{3}{2\pi}\,
 \frac{l-1}{2l+1}\frac{e_0}{r_0}\, Z^2\, A^{-1/3} \;,
\end{equation}
which are diagonal and independent of $m$ like the mass tensor.

\section{Results on stability and instability}\label{results}

According to (\ref{gf13}) the eigenmode frequencies are obtained by 
a numerical diagonalization of the matrix $B^{-1}C$.
We present detailed results on stability and instability of bulk and 
surface modes as functions of $\varrho$ and $T$.
The calculations have been performed with the Skyrme energy densities,
SkM$^*$ and SIII (cf.\ Table \ref{Skyf}), 
implying soft and stiff equations of state (EOS), respectively.
We present here mainly the results for the soft equation of state
(SkM$^*$), because it is favored {\it e.g.} by the study of monopole
vibrations and supernova explosions. 

\begin{figure}
\begin{center}
\resizebox{0.48\textwidth}{!}{%
  \includegraphics{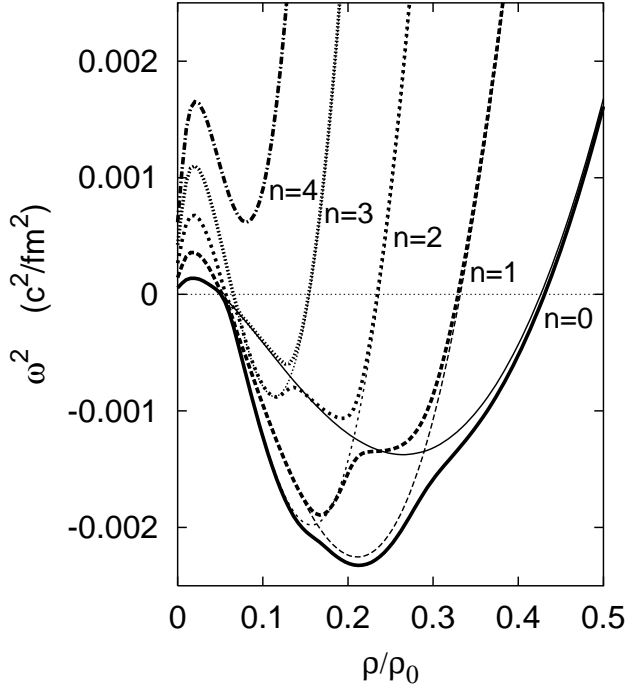} }  \end{center}
\caption{Eigenvalues $\omega^2$ for different compressional modes of quadrupole  
oscillations $(l=2)$ as functions of density at temperature $T=3$ MeV.
Results for a gold-like droplet described by a soft EOS (SkM$^*$) are
presented. The diagonal contributions are displayed
as thin lines, while the heavy lines represent the results obtained by
diagonalization.}
\label{l2an1-5T3}       
\end{figure}
\begin{figure}
\begin{center}
\resizebox{0.5\textwidth}{!}{%
  \includegraphics{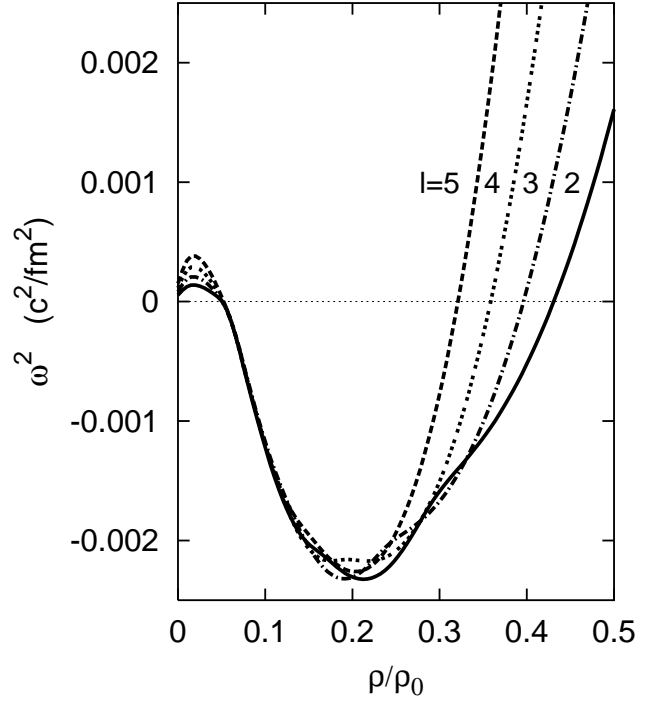} } \end{center}
\caption{Lowest eigenvalues $\omega^2$ for the multipoles
$l=2,3,4,5$ as functions of density at temperature $T=3$ MeV.    
Results for a gold-like droplet are presented for the soft EOS.
Note that the line for $l=2$ is the lowest eigenvalue $\omega^2$
shown as heavy line in fig.\ \ref{l2an1-5T3}}
\label{l2-5mod}	 
\end{figure}

\subsection{Compressional (bulk) modes}\label{Au}

\begin{figure}
\begin{center}
\resizebox{0.48\textwidth}{!}{%
  \includegraphics{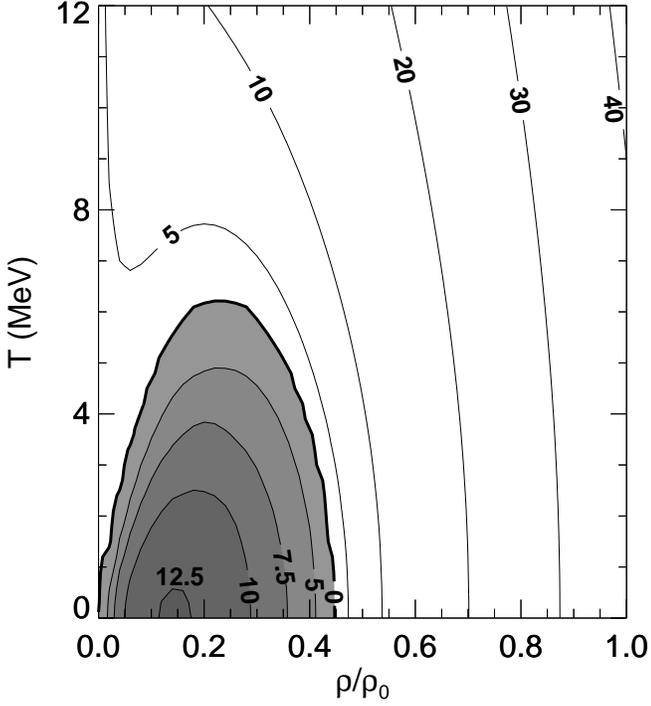}
}  \end{center}
\caption{Lowest vibrational energies $\hbar\omega$ (in MeV) in the stable
(white) region and largest growth rate $\gamma=\hbar\,\sqrt{-\omega^2}$
(in MeV) in the unstable (shaded) region
as functions of $\varrho$ and $T$ for
a gold-like droplet  described by a soft EOS (SkM$^*$).
All modes with $l=2,3,4,5$ and $n=0,1,2,3,4,5$ are taken into account.
To obtain the growth time $\hbar/\gamma$ in fm/c one has to divide 
197 by the $\gamma$ value given in the shaded area.}
\label{tl2-5soft}       
\end{figure}

The results on compressional modes are presented in 
figs.\ \ref{l2finsiz}--\ref{tl2-5soft}.

\subsubsection{Finite-size effects}\label{Fse}

In fig.\ \ref{l2finsiz} we illustrate effects of finite size on the 
region of instability in the ($\varrho,T$)-plane, where  
$\varrho$ and $T$
denote the density and temperature of the homogeneous droplet.
As discussed in sect.\ \ref{infsssc}, the spinodal line
for infinite nuclear matter is determined by 
$\mathrm{d}^2\overline{\epsilon}/\mathrm{d}\varrho^2=0$  
and is shown by the heavy solid line. 
For finite nuclei the spinodal regime is considerably 
reduced as shown for the gold- and tin-like nuclear droplets with 
$Z,A=79,197$ and $50,100$, respectively. The spinodal 
line is determined by the $l=2$ modes, as will be discussed below. 

\subsubsection{Main contributions}\label{maincon}

Figure \ref{l2CBa} shows quantitatively the importance of different
contributions to the lowest eigenvalue $\omega^2$ as function 
of the density for $l=2$ at a temperature of 4 MeV.
The eigenvalues are determined essentially only by the terms 
$C^{(\tau)}, C^{(V)}$ and $C^{(W)}$, which are due to the intrinsic
kinetic energy and the nuclear interaction parts $\epsilon_V$ 
and $\epsilon_W$. The contributions from Coulomb interactions 
and surface energy are negligible, which is well known for
compression modes \cite{b43}. Figure \ref{l2CBa} also reveals 
that the additional Weizs\"acker contribution is the main reason
for the reduction of the spinodal region in the finite systems, 
cf.\ the discussion at the end of sect.~\ref{infsssc}.
Since $\kappa_{nl}=0$ on the spinodal line of an infinite system 
$(R\rightarrow\infty)$, the Weizs\"acker term does not contribute.
The difference between the spinodal lines for $A=100$ and $A=197$ is 
only marginal, because $R \propto A^{1/3}$ and 
$\kappa_{nl} \propto A^{-1/3}$.

\begin{figure}[th]
\begin{center}
\resizebox{0.48\textwidth}{!}{%
  \includegraphics{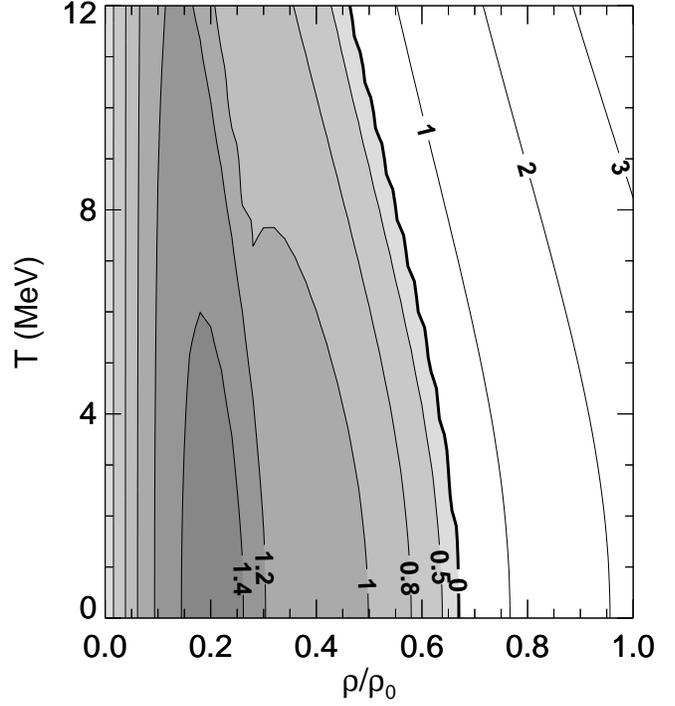} }
\end{center}
\caption{Contour plot of stability and instability for the
surface modes of a gold-like droplet in the $(\varrho,T)$-plane.
Conventions of presentation are the same as in fig.~\ref{tl2-5soft}.}
\label{tl2-5BM}       
\end{figure}

Figure \ref{l2CBahard} illustrates how the contributions to $\omega^2$
change for the stiff EOS.
The results are similar to those for the soft EOS,
but with larger instability inside the spinodal region and larger
stability outside. 

\subsubsection{Dependence on $n$ and $l$}\label{Dnl}

With decreasing densities and temperatures more and mo\-re modes 
with $n>0$ become unstable for the same multipolarity $l$. 
This feature is illustrated in fig.~\ref{l2an1-5T3}, 
where the eigenvalues $\omega^2$  
of the lowest modes for a gold-like system
are displayed as functions of $\varrho$ at $T=3$ MeV. 
Effects of coupling between different $n$-modes are clearly seen 
near the crossings of the diagonal contributions. Such a behavior is
typical for all multipole modes as well as for the soft and stiff EOS.

Figure~\ref{l2-5mod} illustrates the lowest eigenvalues $\omega^2$ 
for different multipolarities $(l=2,3,4,5)$.
With decreasing density below $0.3\varrho_0$ 
all eigenvalues for the different $l$-values become degenerate.
This means that all multipole distortions have the same growth rate.
Note that also the eigenvalues for different $n$-values become 
degenerate in this range of densities (cf.~\cite{b26,b25}). 
As is discussed in sect.\
\ref{effbul}, such a feature can be responsible for a power law in the
fragment-mass distribution of multifragmentation. 

\subsubsection{Dependence on $\varrho$ and $T$}\label{DroT}

Figure \ref{tl2-5soft} shows the lowest eigenvalues $\omega^2$
as functions of $\varrho$ and $T$ for a gold-like droplet 
described by a soft EOS (SkM$^*$). For convenience the numbers 
on the contour lines are not the $\omega^2$-values but
the vibration energies in the stable region
and the growth rates in the unstable region. The stable
vibration $\hbar\omega=28$ MeV of gold at $T=0$
is considerably larger than
the quadrupole energy (20.6 MeV) given  in \cite{b43}, the difference
being due to the additional Weizs\"acker term in our calculations.
The spinodal line for $A=197$ 
of fig.~\ref{l2finsiz} is identical with the 0-line
in fig.~\ref{tl2-5soft}. We note in passing that the inclusion of 
$l=1$ and $l=0$ modes gives only a marginal increase of the region
of instability. 
 
\subsection{Surface modes}\label{surface}

\begin{figure}[th]
\begin{center}
\resizebox{0.48\textwidth}{!}{%
  \includegraphics{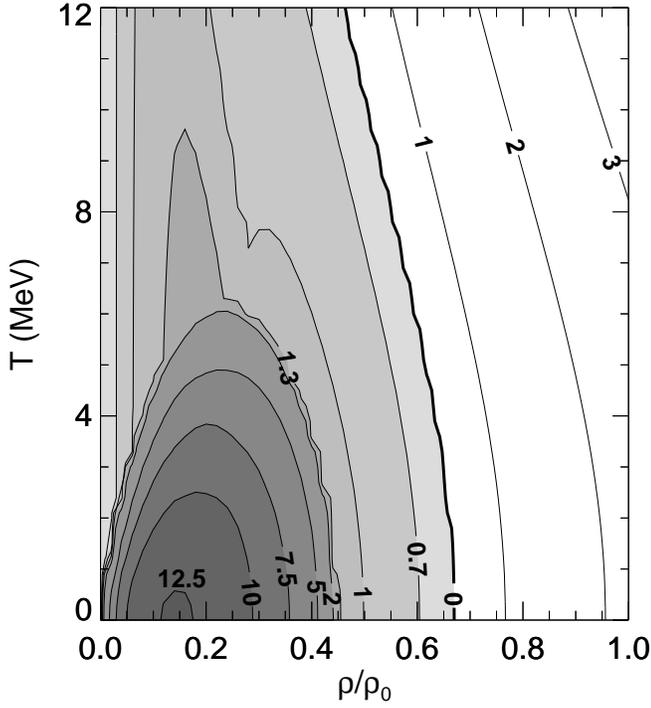} }
\end{center}
\caption{Combined bulk and surface instabilities (soft EOS) for a
gold-like droplet.
Shown are the largest growth rates and the lowest vibrational energies.
Conventions of presentation are the same as in fig.~\ref{tl2-5soft}.}
\label{tl2-5busu}       
\end{figure}

The results are summarized in fig.~\ref{tl2-5BM}. In contradistinction
to fig.~\ref{tl2-5soft} for the compression modes the borderline 
between stability and instability for the surface modes  
is located at considerably larger densities near the spinodal line
of infinite nuclear matter (by accident probably) and extends
to high temperatures. This behavior is due to the softening of the
surface tension with decreasing density and the weak 
temperature dependence.
Thus, there is a wide region for large densities and temperatures, 
where the system becomes unstable with respect to surface deformations 
while being stable for compressional modes. However, the magnitudes
of growth rates and vibrational energies of the surface modes
are smaller by about half an order of magnitude as compared to the
values for the compression modes.
Note that the quadrupole mode is by far most unstable as seen from the  
$l$-dependence of the stiffness tensors (\ref{SstenBM}) and 
(\ref{CstenBM}). 

\begin{figure}[th]
\begin{center}
\resizebox{0.48\textwidth}{!}{%
  \includegraphics{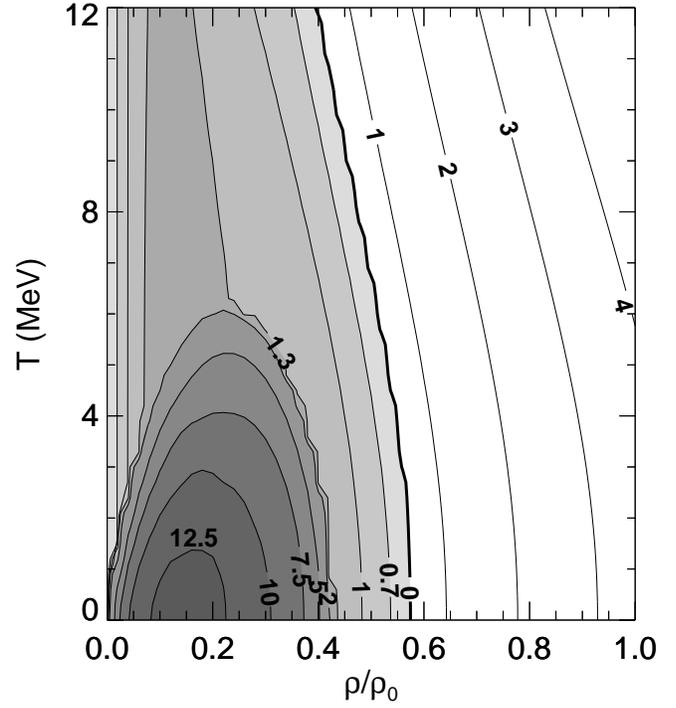} }
\end{center}
\caption{The same as in fig.~\ref{tl2-5busu}, but for tin-like droplet.}
\label{tl2-5NZbusu}       
\end{figure}

\begin{figure}[tbh]
\begin{center}
\resizebox{0.48\textwidth}{!}{%
  \includegraphics{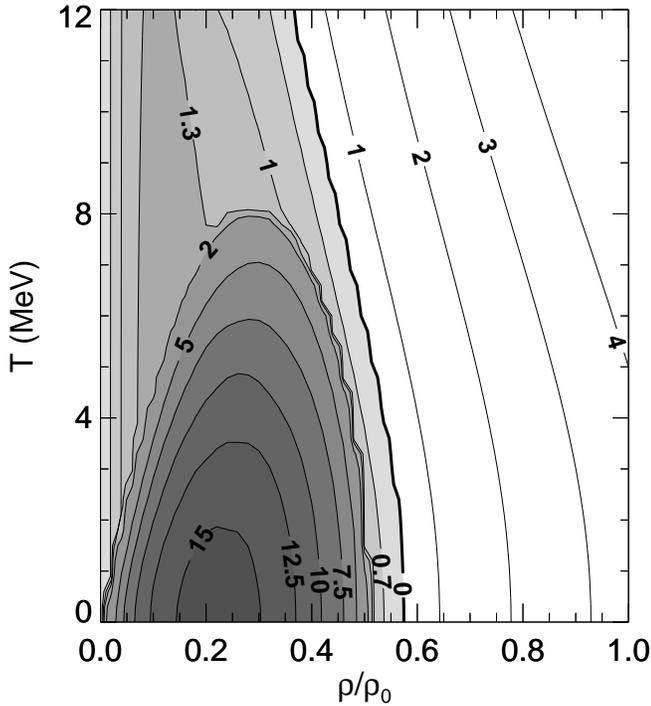} }
\end{center}
\caption{The same as in fig.~\ref{tl2-5busu} but for a stiff EOS .}
\label{tl2-5busuh}       
\end{figure}

\subsection{Combined bulk and surface modes}\label{cbasm}

A combined contour plot of compression and surface instabilities 
is presented in fig.~\ref{tl2-5busu}. Compression 
instabilities are dominant at small densities and
temperatures, while for large temperatures and densities 
only the surface modes are unstable. 

\subsubsection{Dependence on size}\label{dos}

Figure \ref{tl2-5NZbusu} illustrates the regions of stability and
instability for a tin-like droplet. As already shown in 
fig.~\ref{l2finsiz}, the compression instabilities of finite nuclear 
droplets depend only weakly on their masses $A$.
The depth of the compression instability hole (dark regions) 
remains almost the same for the gold- and tin-like droplets.
However, due to the sensitive balance between Coulomb and surface
energies, the surface instabilities are pushed towards  
smaller densities for the lighter system.

\subsubsection{Dependence on EOS}\label{doEOS}

For the same mass of the droplet the regions of instability 
depend on the EOS.
Figure~\ref{tl2-5busuh} displays
the combined instabilities for the gold-like droplet and a stiff EOS.
Comparing the plot with fig.~\ref{tl2-5busu} we realize that the 
compression instabilities extend to higher temperatures for the 
stiff EOS (up to 8 MeV, as compared to about 6 MeV for the soft EOS) 
and to somewhat larger densities. In addition, the depth of the 
compression instability hole is substantially larger for the stiff EOS. 
The values for the growth rates reported for $\varrho=0.37\varrho_0$,
$T=0$ in \cite{b26},  $\varrho=0.3\varrho_0$, $T=3$ MeV in \cite{b27}
and $\varrho=0.3\varrho_0$, $T=6$ MeV in \cite{b24} are
consistent with the values in the contour plot of fig.~\ref{tl2-5busuh}.

\section{Instabilities and multifragmentation}
\label{instmul}

As compared to infinite nuclear matter the spinodal region of the bulk
(compression) instabilities is significantly reduced in finite systems
(cf.\ fig.~\ref{l2finsiz}). However, finite systems exhibit additional
modes due to the surface. Surface instabilities arise in finite systems
already below $\varrho/\varrho_0\approx 0.6$ and extend to high 
excitation
energies of the system (cf.\ fig.~\ref{tl2-5BM}) for $A\approx 200$.
Therefore in addition to bulk instabilities, surface instabilities 
become important in multifragmentation reactions.
A heated nuclear droplet expands essentially at constant entropy 
(cf.\ fig.~2, \cite{b45}), and hence experiences, with increasing
initial excitation energy, first surface instabilities and then 
bulk instabilities. When reaching the spinodal regime the temperatures  
are of the order of about 5 MeV. In the following
we discuss possible effects from surface and bulk instabilities, which   
become important during the expansion of the hot pieces of nuclear matter
formed initially in fragmentation reactions.

\subsection{Effects from surface instabilities}
\label{effsur}

Hot spherical nuclei are formed initially most likely in high-energy 
target and projectile fragmentation reactions, as well as in 
bombardments of heavy nuclei with light particles 
$(p,d,\alpha,\ldots)$. If such hot nuclei are formed with 
excitation energies in the range 2.5 MeV $\lesssim E^*/A 
\lesssim 5$ MeV (6 MeV$\lesssim T \lesssim 9$ MeV) we expect \cite{b45} 
expansion into spinodal region of surface instabilities.
Since quadrupole deformations $(l=2)$ are by far the most unstable 
surface modes (cf.\ eqs.\ (\ref{SstenBM}),(\ref{CstenBM})), partition 
into
two fragments of almost equal size ($A_f \approx A/2$) is most likely.
This type of fragmentation differs qualitatively from the normal fission
mechanism, because it happens at low densities, where no barrier is 
present any more.

In central collisions of (almost) equal nuclei at bombarding energies 
$\gtrsim$
100 MeV/u one expects large dynamical distortions of the nuclear matter  
\cite{b46}. Because of the large times
which are necessary to restore spherical symmetry (according to 
fig.~\ref{tl2-5BM} these times are of the order of (100-200) fm/c), 
the deformed hot matter will expand (typical time 30 fm/c) 
into the instability regime and fragment by pure surface instability 
plus dynamical collective motion or by additional bulk instabilities.

\subsection{Effects from bulk (compressional) instabilities}
\label{effbul}

For hot spherical nuclei formed at excitation energies 
$E^*/A$ $\gtrsim 5$ MeV
($T\gtrsim 9$ MeV) we expect \cite{b45} expansion into the
spinodal regime of bulk instabilities. Near the spinodal line (of bulk
instabilities, cf.\ fig.~\ref{l2-5mod}) quadrupole instabilities are most
likely to develop. However, as one proceeds in reaching smaller densities 
modes with larger $l$-values and also larger $n$-values (cf.\ 
fig.~\ref{l2an1-5T3} and \ref{l2-5mod}) become unstable, reaching equal
growth times for $\varrho/\varrho_0 \lesssim 0.25$.

This feature leads naturally to a power law in the fragment mass 
distribution. In an instability mode, which leads to 
$m$ fragments, one produces fragments with mass $A_f = A/m$ 
with probability 
\begin{equation} \label{efbul1}
P_m(A/m) \propto m^2 \;,
\end{equation}
where the second factor $m$ is due to the degeneracy of this mode.
Since $m= A/A_f$  we find 
\begin{equation} \label{efbul2}
P(A_f) \propto A_f^{-2}
\end{equation}
for the fragment mass distribution. Of course, such a behavior can only
be expected for $m>2$, and hence $A_f$ well below $A/2$.
We stress that the power-law behavior (\ref{efbul2}) is due only to
the degeneracy of growth times for different modes $n,l$, and not related 
to the critical point. 

\subsection{General aspects of fragment-mass distributions}
\label{genasp}

According to expansion and instabilities of spherical drop\-lets 
we expect the following qualitative features for the frag\-ment-mass 
distribution with two or more heavy fragments.
\begin{itemize}
\item For low energies only fission-like fragmentation should occur
which give rise to a bump around $A_f\lesssim A/2$ and emitted light particles 
with an exponentially decreasing mass distribution. 
\item For energies 5 MeV $<E^*/A<$ 8 MeV,
which lead to expansion just into the spinodal region
of bulk instabilities, the fragmentation is governed by low $l$-values
($l=1,2,3$). According to fig.~\ref{modshap} large regions of low densities 
are produced in the development of these low-$l$ instabilities.
The subsequent fragmentation of these will lead to a power-law behavior
for light fragments in addition to the heavy fragments and their 
evaporated light particles.
\item For energies 8 MeV $<E^*/A<$ 12 MeV, 
which lead to expansion well into the region of bulk
instabilities  for densities $\varrho/\varrho_0 < 0.3$, a pure power law 
with $\sigma=2.0$ is expected for the fragment mass distribution.
\end{itemize}

These properties of fragment-mass distributions are well established  
by experiments, cf.\ \cite{b38}. The observed larger values of $\sigma$ 
for $E^*/A>$ 12 MeV indicate that the freeze-out of fragments 
is probably faster than allowed by growth rates of instabilities, 
{\it e.g.} by a coalescence mechanism.

\subsection{Experimental prospects}\label{conclusions}

It is not clear how those components of the multifragmentation process 
can be identified, which are due to spinodal decomposition. 
A possible strategy is to analyze the data event by event and to look 
for characteristic modes as function of initial excitation energy 
\cite{b27,b47}. An indication for spinodal decomposition
has been reported recently \cite{b28}.

Of course, one should look particularly into fragmentation reactions, 
which are favorable for spinodal decomposition occurring in large 
dilute spherical droplets. As mentioned earlier in sect.\ \ref{effsur}, 
central heavy-ion collisions exhibit violent collective deformations, 
and hence will have complicated dynamics,
which may mask the spinodal decomposition phenomena. 

In projectile and target fragmentation at high 
bombarding energies the excitation by abrasion yields
excited spectators without large dynamical distortion. Thus we can
expect that a large fraction in the multifragmentation process 
is due to spinodal decomposition after expansion. 
However, induced by the abrasion process, certain
parts of projectile and target will be emitted initially 
before expansion. This component, which may contribute of order 10\% 
to the fragment yield, is characterized by considerably larger fragment
kinetic energies. During the fast expansion the number of emitted particles 
is small \cite{b45}, such that we expect only two relevant components,
one from the initial excitation process and another one from the break-up 
after expansion. 

A recent analysis \cite{b48} of the decay of target spectators in 
$^{197}$Au+$^{197}$Au collisions at 1 GeV/u has indicated that the fragments 
are already produced at the initial matter densities. However, we would
like to interpret the observed fragment kinetic energy spectra as evidence 
for a high- and low-energy component, which correspond to initial 
excitation process and the break-up after expansion, respectively. 
Such a picture is supported by the decrease of the He-Li temperatures from
(10--12) MeV in the high-energy tails to about 5 MeV in the low-energy part 
of the fragment kinetic-energy distribution. The evolutionary character 
of the multifragmentation process from initial excitation via expansion to the
final break-up is supported by the analysis of central 
$^{129}$Xe+$^{\mathrm{nat}}$Cu at 30 MeV/u \cite{b54*} and light-ion induced 
fragmentation reactions \cite{b52*}.

\begin{acknowledgement}
\subsection*{Acknowledgements}
We gratefully acknowledge fruitful discussions with our experimental 
colleagues 
U.~Lynen, W.~Trautmann and C.~Schwarz.
\end{acknowledgement}

\begin{appendix}
\section*{Appendix}
\section{The displacement transformation}\label{ap0}

Displacement transformations have been frequently used in the
description of small-amplitude collective nuclear motion 
\cite{b49,b50,b51,b52}. With the introduction of the displacement
for wave functions in a differential form it was possible to
extend such descriptions to large-amplitude dissipative collective
motion, cf.\ sect.~\ref{reminder}. 
Within this approach a scaling condition (\ref{filam})
for the stationary part of the diabatic single-particle wave functions,
written here as
\begin{equation}\label{aa1}
 \frac{\partial}{\partial t}
 \widetilde{\phi}(\vec{r},t) = -\frac{1}{2}[\vec{v}
 (\vec{r},t)\cdot\vgrad +\vgrad\cdot \vec{v}(\vec{r},t)]\, 
 \widetilde{\phi}(\vec{r},t) \;,
\end{equation}
assures that all dynamical couplings linear in the velocities
disappear in the Schr\"odinger equation for the time-dependent wave 
function. 
Considering this equation (\ref{aa1}) at the time dependent point 
$\vec{r'}=\vec{r}+\vec{s}(\vec{r},t)$
we find with $\dot{\vec{s}}(\vec{r})=\vec{v}(\vec{r}+\vec{s}(\vec{r}))$
for the total derivative
\begin{equation}\label{aa2}
 \frac{\rmd}{\rmd t} \widetilde{\phi}(\vec{r}+\vec{s}(\vec{r},t),t) =
 \widetilde{\phi}(\vec{r}+\vec{s}(\vec{r},t),t)\,(-\frac{1}{2}\vgrad'\cdot
 \dot{\vec{s}}(\vec{r}))\;,
\end{equation}
where $\vgrad'$ denotes the gradient vector taken 
at the position $\vec{r'}$.
After integration from $t=0, s=0$ to $t, \vec{s}(t)$
\begin{equation}\label{aa3}
 \widetilde{\phi}(\vec{r}+\vec{s}(\vec{r},t),t) 
  =  \phi(\vec{r})\exp\{-\frac{1}{2}
 \vgrad'\cdot\vec{s}(\vec{r},t)\} \;, 
\end{equation}
{\it i.e.} the value of the transformed wave function at the point 
$\vec{r'}=\vec{r}+\vec{s}(\vec{r},t)$ is given by the original
wave function $\phi(\vec{r})$
at the point $\vec{r}$ normalized to ensure
the unitarity of the transformation as implied by the integrated 
eq.\ (\ref{aa1}),
\begin{equation}\label{aa4}
 \widetilde{\phi}(\vec{r'},t) = 
 \exp\{ -\frac{1}{2}[\vec{s}(\vec{r},t)\cdot\vgrad'
  + \vgrad'\cdot\vec{s}(\vec{r},t) ] \} \,\widetilde{\phi}(\vec{r'},0) \;.
\end{equation}
Recently, ~similar unitary displacement transformations have been
successfully implemented in the treatment of two-body correlations 
caused by realistic two-body interactions \cite{b53}.

\subsection{Continuity equation, particle conservation}\label{ap1}

Defining the total single-particle density by the sum over all
occupied states, we find as a consequence
of (\ref{aa3}) 
\begin{equation}\label{aa5}
 \widetilde{\varrho}(\vec{r}+\vec{s}(\vec{r},t),t) =
 \varrho (\vec{r}) \,\rme^{-(\vgrad'\cdot \vec{s}(\vec{r},t))}\,,
\end{equation}
where $\varrho(\vec{r})$ is the density of the undistorted sphere.
In differential form this equation becomes (cf.~(\ref{aa2}))
\begin{equation}\label{aa6}
 \frac{\rmd \widetilde{\varrho}(\vec{r'},t)}{\rmd t} = 
 - \widetilde{\varrho}(\vec{r'},t) (\vgrad'\cdot \dot{\vec{s}}(\vec{r},t))
\end{equation}
and, since 
$\dot{\vec{s}}(\vec{r},t)= \vec{v}(\vec{r}+\vec{s}(\vec{r}),t)$,
we find the continuity equation 
\begin{equation}\label{aa7}
 \frac{\rmd \widetilde{\varrho}(\vec{r'},t)}{\rmd t} + 
 \widetilde{\varrho} (\vec{r'},t)\,
 (\vgrad\cdot \vec{v})_{\vec{r'}} = 0\;,
\end{equation}
which also results form (\ref{gf10}) by considering
the substantial change of density $\rmd \widetilde{\varrho}(\vec{r'})/
\rmd t$ along the displacement path, {\it i.e.} in the comoving frame.

Particle conservation implies 
\begin{equation}\label{aa7p}
 \rmd^3 r'\,\widetilde{\varrho}(\vec{r'}) = \rmd^3 r\,\varrho(\vec{r}) \;,
\end{equation}
which means, that the number of particles in a volume
element is conserved when followed in the distortion process.

\subsection{Conservation of the center of mass }\label{ap2}

The time derivative of the center of mass
\begin{equation}\label{cm0}
 \frac{\mathrm{d}}{\mathrm{d} t} \langle \vec{r}\rangle =
 \frac{1}{A} \int \mathrm{d}^3 r'\, \widetilde{\varrho}(\vec{r'})\,
 \vec{v}(\vec{r'},t)
\end{equation}
should vanish. This integral is rewritten as
\begin{equation}\label{cm1}
\frac{\mathrm{d}}{\mathrm{d} t}  \langle \vec{r}\rangle = 
\frac{1}{A} \int \mathrm{d}^3 r\, 
\varrho (\vec{r})\,\dot{\vec{s}}(\vec{r},t) \;,
\end{equation}
by using (\ref{aa7p}) and 
$\vec{v}(\vec{r'},t)= \dot{\vec{s}}(\vec{r},t)$
with $A$ and $\varrho(\vec{r})$  denoting the mass number and the unperturbed 
density distribution, respectively. 
This equation is understood by noting 
that the mass element positioned at $\vec{r}$
for $\vec{s}=0$ contributes for $\vec{s}\neq 0$ at the point \
$\vec{r}+\vec{s}(\vec{r},t)$ to the center of mass.
Since $\vec{s}$ is given by (\ref{gf8}) we find by partial 
integration of (\ref{cm1}) that
\begin{equation}\label{cm2}
\frac{\mathrm{d}}{\mathrm{d} t}
\langle \vec{r}\rangle \propto \sum_\lambda \dot{q}_\lambda \,
\chi_\lambda(R,\Omega) = 0 \, ,
\end{equation}
{\it i.e.} for every multipole distortion (even for $l=1$) the center of mass
is conserved exactly.

\subsection{Second-order expansion of density}\label{ap3}

Expanding (\ref{aa5}) to second order in $\vec{s}$, we find
\begin{equation}\label{a0}
 \widetilde{\varrho}(\vec{r}+\vec{s},t)  = \varrho (\vec{r})\,
 [ ( 1- \vgrad\cdot\vec{s}+ \frac{1}{2}(\vgrad\cdot\vec{s})^2 
  -(\vgrad'-\vgrad)\cdot \vec{s}] \;.
\end{equation}
The last term can be rewritten in second order as 
\begin{equation}\label{a0a}
-(\vgrad'-\vgrad)\cdot \vec{s} = 
 \frac{1}{2}\sum_{i,j}(\partial_i\partial_j\, w)^2 \;,
\end{equation}
where $\partial_i$ denotes the differentiation with respect to
the $i$'th cartesian component of $\vec{r}$. 
From (\ref{gf5}),(\ref{gf7}) and (\ref{gf8}) we obtain 
\begin{equation}\label{a01}
-\vgrad\cdot\vec{s} = - \sum_{\lambda}\,q_{\lambda}\Delta\chi _{\lambda}
=  \sum_{\lambda}\,q_{\lambda}\, \kappa_{nl}^2\,\chi _{\lambda} \;.
\end{equation}
The integral $\int_{r\leq R}\,\mathrm{d}^3 r\, \vgrad\cdot\vec{s}=0$,
because the integral over the single spherical harmonic in 
$\chi_\lambda$ vanishes for $l>0$. This property causes 
the first-order derivative of the total energy of the spherical droplet
to vanish, and hence the spherical droplet is an equilibrium point with
respect to the displacement field.  

 For the derivatives of the density 
with respect to the collective coordinates $q_\lambda$
we obtain the relations
\begin{equation}\label{a02}
 \left (\frac{\partial \widetilde{\varrho}}{\partial q_\lambda}
 \right )_{\vec{q}=0} = \varrho(\vec{r})\, \kappa_{nl}^2\,\chi_{\lambda}
(\vec{r})
\end{equation}
and 
\begin{eqnarray}\label{a03}
 \left( \frac{\partial^2 \widetilde{\varrho}}
 {\partial q_\lambda \partial q_{\lambda'}}
 \right)_{\vec{q}=0} 
 & = &\varrho(\vec{r})\, [ 
 \kappa_{nl}^2 \kappa_{n'l'}^2\,\chi _{\lambda}
 (\vec{r}) \chi _{\lambda'}(\vec{r})  \\ 
 & &  +  \sum_{i,j} (\partial_i\partial_j
 \chi _{\lambda})(\partial_i\partial_j\chi _{\lambda'}) 
 ] \;, \nonumber 
\end{eqnarray}
which frequently appear in the evaluation of the stiffness tensor
(cf.\ Appendix \ref{aa}).    
  
\section{Surface tension}\label{bb} 

The surface interaction energy $E^{LD}_{S}$ of a sphere with radius
$R$ is determined by the difference 
\begin{equation} \label{esld} 
E^{LD}_{S} = E_V + E_W - E^{LD}_{V}
\end{equation}
of the local and nonlocal interaction energies $E_V+E_W$
for the diffusive sphere and the interaction energy $E^{LD}_{V}$ 
of the homogeneous sphere with the constant density 
$\varrho$ in the interior $(r<R)$.

Assuming a linear approximation 
\begin{equation} \label{rolin} 
 \widetilde{\varrho}(r,t) = \frac{1}{2}\varrho \left(1-\frac{r-R}{2a}\right)
\quad \mbox{for} \quad |r-R| \leq 2a 
\end{equation}
to the Fermi function for describing the transition of the density
from the inside to the outside ($a$ denoting the diffuseness
parameter) we obtain for $R\gg a$ (leptodermous system)
\begin{equation} \label{delev}
E_V-E^{LD}_{V} = -4\pi R^2\,\frac{2}{3} e_V \varrho^2 a\;,
\end{equation} 
\begin{equation} \label{delew} 
E_W = 4\pi R^2\,\frac{1}{4} e_W \frac{\varrho^2}{a}\;. 
\end{equation} 
Here, the energy densities $\epsilon_V= e_V \varrho^2(r,t)$ and 
$\epsilon_W= e_W (\vgrad\varrho)^2$ have been used.
Dividing by the surface $4\pi R^2$, we find the surface tension
\begin{equation} \label{epss}
\epsilon_S= \varrho^2\left(-\frac{2}{3}e_V a(T)+\frac{1}{4}
\frac{e_W}{a(T)}\right) \;,
\end{equation}
where $a(T)$ increases with increasing temperature. 
According to \cite{b54} $a(T)/a(0)-1\propto T^2$, and hence 
\begin{equation} \label{epssT} 
\epsilon_S(\varrho,T) =  \epsilon_S(\varrho_\mathrm{eq},0)\, 
\left( \frac{\varrho}{\varrho_\mathrm{eq}}\right)^2 (1+\beta T^2) \;.
\end{equation}
Introducing the explicite form for $\epsilon_S(\varrho_\mathrm{eq},0)$
with the asymmetry dependence on neutron and proton numbers, we finally
obtain eq.~(\ref{gf21}) with the values $\beta$ from 
\cite{b41,b42}.

\section{Evaluation of the stiffness tensor}\label{aa} 

The evaluation of the different contributions to the stiffness tensor
(\ref{gf22})
is simplified by considering the displacement of individual
volume elements with the displacement field $\vec{s}(\vec{r})$.
Then according to (\ref{aa7p}),
the integrals over the distorted sphere can be replaced
by integrals over the original volume elements. 
In this way only integrals over the original sphere of
radius $R$ have to be evaluated.

\subsection{Intrinsic kinetic energy contribution $\Cllt$}\label{aaint}

In the local-density approximation the intrinsic kinetic energy 
for protons or neutrons is given by
\begin{equation}\label{aint1}
 E_{\mathrm{kin}}^{\mathrm{intr}} = 
 \int_{r\leq R}\,\mathrm{d}^3 r \frac{\hbar^2}
 {2m^*(\widetilde{\varrho})} \,\int \mathrm{d}^3 k \, 
 f(k)\widetilde{\vec{k}}^2 \;,
\end{equation}
where $f(k)$ denotes the (isotropic)
momentum distribution within the unperturbed droplet
normalized to $\varrho$.
Due to the distortion, the local momentum $\vec{k}$ 
transforms to $\widetilde{\vec{k}}$.
In the {\em adiabatic} limit the
density dependence of $\langle \widetilde{\vec{k}}^2 \rangle
\propto \widetilde{\varrho}^{2/3}$ enters, and hence with the 
density (\ref{a0}) at the displaced volume element we obtain
\begin{equation}\label{aint6}
 \langle \widetilde{\vec{k}}^2 \rangle = \langle \vec{k}^2 \rangle
 \{ 1 -\frac{2}{3}\vgrad\cdot\vec{s}
 +\frac{7}{18}(\vgrad\cdot\vec{s})^2 + \frac{1}{3} \sum_{i,j}
 (\partial_i \partial_j\, w)^2 \}
\end{equation}
up to second order in the displacement field $\vec{s}$.

In the final evaluation of 
\begin{eqnarray}\label{aint7}
 \Cllt & =&  \left( \frac{\partial^2\, E_{\mathrm{kin}}^{\mathrm{intr}}}
 {\partial q_\lambda\partial q_{\lambda'}}
 \right)_{\vec{q}=0} \\
 & = & \varrho \int
 \mathrm{d}^3 r \frac{\hbar^2}{2} \left \{ \frac{\partial^2}
 {\partial q_\lambda\partial q_{\lambda'}} 
 \frac{\langle \widetilde{\vec{k}}^2 \rangle}{m^*(\widetilde{\varrho})}
 \right \}_{\vec{q}=0} \nonumber
\end{eqnarray}
we encounter integrals 
\begin{equation}\label{aint8}
 \int\mathrm{d}^3 r (\vgrad\cdot\vec{s})^2 =
 \sum_{\lambda} \,q_\lambda^2\, \kappa_{nl}^4
\end{equation}
and
\begin{eqnarray}\label{aint9}
 \sum_{i,j} \,\int\mathrm{d}^3 r\,(\partial_i\partial_j
\chi _{\lambda})(\partial_i\partial_j\chi _{\lambda'}) 
 &=& \hspace{20ex} \\
\int \mathrm{d}^3 r\, \{ \sum_{i,j}
\partial_i[(\partial_j \chi _{\lambda})
(\partial_i\partial_j\chi _{\lambda'})]  &-& 
	(\partial_j \chi _{\lambda})
\partial_i\partial_i (\partial_j\chi_{\lambda'})\} \nonumber \,.
\end{eqnarray}
Here, starting with (\ref{aint7}), and in the following we imply
that integration over $\vec{r}$ is limited to the unperturbed sphere
$(r\leq R)$.
In (\ref{aint9}) the second term on the r.h.s.\ is evaluated using 
$\sum_i \partial_i\partial_i =\Delta$ and (\ref{gf5}) twice  
(on the way one more integration by parts and transformation  
to a surface integral that vanishes as $\chi_{\lambda}(R)=0$), which leads to
\begin{equation}\label{a7}
\int \mathrm{d}^3 r\,(\partial_j \chi _{\lambda})
\partial_i\partial_i (\partial_j\chi_{\lambda'}) =
\kappa_{nl}^2\kappa_{n'l'}^2\, \delta_{\lambda\lambda'} \,.
\end{equation}
The first term on the r.h.s.\ of (\ref{aint9}) is 
transformed to the symmetric form
\begin{eqnarray}\label{a8}
 \sum_{i,j} \int \mathrm{d}^3 r\, \partial_i[(\partial_j \chi _{\lambda})
(\partial_i\partial_j\chi _{\lambda'})] &=& \hspace*{30mm}  \\ 
\frac{1}{2} \sum_{i,j} \int \mathrm{d}^3 r\,\partial_i\partial_i
(\partial_j \chi _{\lambda}\partial_j \chi _{\lambda'}) & = &
  - \frac{4 \kappa_{nl}\kappa_{n'l'}}{R^2}\, \delta_{ll'}\delta_{mm'} 
 \nonumber \,.
\end{eqnarray}
The latter result is obtained by
transforming to a surface integral and utilizing the 
properties of derivatives of Bessel functions on the surface,
\begin{equation}\label{bessur}
j''_l(\kappa_{nl}R) = - \frac{2}{\kappa_{nl}R}\, j'_l(\kappa_{nl}R)\;,
\end{equation}
which result from their differential equation taken at \\$r=R$.

The final result given by eq.~(\ref{gf23})
is obtained by putting all terms together, including the linear
density dependence of $(m^*(\widetilde{\varrho}))^{-1}$ and 
summing over protons and neutrons.

\subsection{Interaction-energy contribution $\CllV$}\label{bien}

 This term is calculated from the expression
\begin{equation}\label{bien1}
 \CllV = \left\{ \frac{\partial^2 }{\partial \ql\partial \qlp}
 \int\mathrm{d}^3 r'\,\epsilon_V(\widetilde{\varrho}(\vec{r}'))
 \right\}_{\vec{q}=0} \;.
\end{equation}
With the explicit introduction of the displacement field 
$\vec{s}(\vec{r})$
we can transform the integral over $\vec{r}'$ to an integral
over $\vec{r}=\vec{r}'-\vec{s}(\vec{r},t)$ by replacing 
$\mathrm{d}^3 r'$ by $\mathrm{d}^3 r\, \varrho(\vec{r})/
\widetilde{\varrho}(\vec{r}')$. We find 
\begin{eqnarray}\label{bien2}
 \CllV &=& 2\left(\frac{1}{2}\frac{\partial^2\epsilon_V(\varrho)}
{\partial\varrho^2}-\frac{1}{\varrho}\frac{\partial\epsilon_V(\varrho)}
{\partial\varrho}+\frac{\epsilon_V(\varrho)}{\varrho^2}\right) 
 \nonumber \\ &&\times
\int\mathrm{d}^3 r\,\left(\frac{\partial\widetilde{\varrho}}
{\partial q_\lambda}\frac{\partial\widetilde{\varrho}}
{\partial q_\lambda'}\right)
_{\vec{q}=0}  \\
&& + \left(\frac{\partial\epsilon_V(\varrho)}{\partial\varrho}-
\frac{\epsilon_V(\varrho)}{\varrho}\right)
\int\mathrm{d}^3 r\,\left(\frac{\partial^2\widetilde{\varrho}}
{\partial q_\lambda\partial q_\lambda'}\right)
_{\vec{q}=0}\nonumber\;,
\end{eqnarray}
where $\varrho(\vec{r})=\varrho$ for $r\leq R$ is used. 
Insertion of (\ref{a02}), (\ref{a03}) and (\ref{a8}) yields the result 
given by (\ref{gf26}).

\subsection{Weizs\"acker-energy contribution $\CllW$}\label{wien}

This term is calculated from the expression 
\begin{equation}\label{wien1}
 \CllW = e_W\left\{ \frac{\partial^2 }{\partial \ql\partial \qlp}
 \int\mathrm{d}^3 r'\, (\vgrad \widetilde{\varrho})^2_{\vec{r'}=
 \vec{r}+\vec{s}} \right\}_{\vec{q}=0} \;,
\end{equation}
where $\epsilon_W=e_W\,(\vgrad \varrho)^2$ has been used.
Note that the surface energy is treated explicitly below, and
hence is excluded here. 
The integral over $r'$ is again replaced by an integral over $r$.
However,  since $\vgrad \widetilde{\varrho}={\cal O}(\ql)$
we only need in second order 
$\partial\widetilde{\varrho}/\partial\ql$ given by
(\ref{a02}) without an additional contribution
from the change of integration variable. 
By partial integration and using 
(\ref{gf3}), (\ref{gf5}) and (\ref{gf6}) we find the result (\ref{gf27}).

\subsection{Surface-energy contribution $\CllS$}\label{sen}

This term is calculated from the expression 
\begin{equation}\label{sen1}
 \CllS = \left\{ \frac{\partial^2 }{\partial \ql\partial \qlp}
 \int \mathrm{d} \widetilde{f}\, \epsilon_s(\widetilde{\varrho})
 \right\}_{\vec{q}=0} \;,
\end{equation}
where the integration runs over the deformed surface
\begin{equation}\label{sen2}
 r(\Omega) = R + s_r(R,\Omega)
\end{equation}
with the radial displacement
\begin{equation}\label{sen3}
 s_r = \sum_\lambda \, \ql \frac{\kappa_{nl}}{R} \sqrt{\frac{2}{R}}
 \, {\cal Y}_l^m(\Omega) \;,
\end{equation}
while the other components of $\vec{s}$ vanish (for $r=R$).
Introducing (cf.~\cite{b43}) 
\begin{equation}\label{sen4}
 \mathrm{d} \widetilde{f} = \mathrm{d}\Omega \, r^2(\Omega)
 \sqrt{1+(\vgrad s_r)^2}
\end{equation}
for the deformed surface element, we are left with an integration over 
$\Omega$. One should note that $\vgrad s_r$ contains only 
angular derivatives.
Furthermore, since $\epsilon_S(\widetilde{\varrho})$ is quadratic 
in $\widetilde{\varrho}$, we have ($\vgrad\cdot\vec{s}=0$ on the surface) 
up to second order in $\vec{s}$
\begin{eqnarray}\label{sen4a}
 \epsilon_S(\widetilde{\varrho},R) 
 &=& \epsilon_S(\varrho)
   \{ 1-2(\vgrad'-\vgrad)\cdot\vec{s} \}_{r=R} \\
 &=& \epsilon_S(\varrho)
 \left\{ 1+2\left(\frac{\partial s_r}{\partial r}\right)^2
 \right\}_{r=R} \; . \nonumber
\end{eqnarray}
Thus we find
\begin{eqnarray}\label{sen5}
 \CllS & =  \epsilon_S(\varrho) & \left\{ 
 \frac{\partial^2}{\partial \ql\partial \qlp} \right. \\
 && \left. \int \mathrm{d}\Omega \{9s_r^2(R) 
 + R^2(\vgrad s_r(R))^2 \} \right\}_{\vec{q}=0} \;, \nonumber 
\end{eqnarray}
where $(\partial s_r/\partial r)_{r=R}=-(2/R)s_r(R,\Omega)$ has
been used (cf.\ eqs.\ (\ref{gf8}), (\ref{bessur}), (\ref{sen3})).
Insertion of $s_r$ according to (\ref{gf8}) and partial integration
of the second term yields the expression (\ref{gf29})
for the surface part of the stiffness tensor.

\subsection{Coulomb-energy contribution $\CllC$}\label{cen}

This term is defined by
\begin{equation}\label{cen1}
 \CllC = \left\{ \frac{\partial^2 }{\partial \ql\partial \qlp}
 \frac{1}{2} \int \mathrm{d}^3 r'_1\mathrm{d}^3 r'_2 
 \frac{\widetilde{\varrho}_p(\vec{r}'_1)\widetilde{\varrho}_p(\vec{r}'_2)}
 {|\vec{r}'_1-\vec{r}'_2|}  \right\}_{\vec{q}=0} \;,
\end{equation}
where $\vec{r}'_1=\vec{r}_1+\vec{s}(\vec{r}_1)$,
$\vec{r}'_2=\vec{r}_2+\vec{s}(\vec{r}_2)$ and 
$\widetilde{\varrho}_p=e_0(Z/A)\widetilde{\varrho}$ denotes the charge
density of the distorted sphere. Replacing the integration
variables by $\vec{r}_1$ and $\vec{r}_2$ and using particle
conservation $\mathrm{d}^3 r'\widetilde{\varrho}(\vec{r'})=
\mathrm{d}^3 r\varrho(\vec{r})$ in the displacement from 
$\vec{r}$ to $\vec{r'}=\vec{r}+\vec{s}(\vec{r})$ we find
\begin{eqnarray}\label{cen2}
 \CllC & = &\frac{1}{2} \left( \frac{Ze_0}{A} \varrho\right)^2 \\
 & \times &
 \left\{ \frac{\partial^2 }{\partial \ql\partial \qlp}
 \int \frac{\mathrm{d}^3 r_1\mathrm{d}^3 r_2}
 {|\vec{r}_1+\vec{s}(\vec{r}_1)-\vec{r}_2-\vec{s}(\vec{r}_2)|} 
 \right\}_{\vec{q}=0} \;, \nonumber
\end{eqnarray}
where the integrals are over $r_1,r_2\leq R$ (the unperturbed sphere).
The terms of the integral, which are of second order in $\vec{s}$
or $\ql$ are given by the sum of $I_1$ and $I_2$.
$I_1$ is defined by 
\begin{equation}\label{cen3}
 I_1 = \int \mathrm{d}^3 r_1\mathrm{d}^3 r_2\,
 [\vec{s}(\vec{r}_1)\cdot \vgrad_1][\vec{s}(\vec{r}_2)\cdot \vgrad_2]
 \frac{1}{|\vec{r}_1-\vec{r}_2|} \;.
\end{equation}
Inserting $\vec{s}(\vec{r}_2)=\vgrad_2 w(\vec{r}_2)$,
integrating by parts over $\vec{r}_2$ (noting $w(R)=0$) and using
$\Delta|\vec{r}_1-\vec{r}_2|^{-1}=-4\pi\delta(\vec{r}_1-\vec{r}_2)$,
we find
\begin{equation}\label{cen3a}
 I_1  =  4\pi \int \mathrm{d}^3 r (\vgrad w)^2 \;.
\end{equation}
$I_2$ is related to $I_1$ by
\begin{eqnarray}\label{cen4}
 I_2 & = & \int \mathrm{d}^3 r_1\,
 \vec{s}(\vec{r}_1)\cdot[\vec{s}(\vec{r}_1)\cdot\vgrad_1]\,\vgrad_1
 \int \frac{\mathrm{d}^3 r_2}{|\vec{r}_1-\vec{r}_2|} \nonumber\\
 & = & -\frac{2\pi}{3}\,\int \mathrm{d}^3 r \,\vec{s}(\vec{r})\cdot
 [\vec{s}(\vec{r})\cdot\vgrad] \vgrad r^2 = -\frac{1}{3}\, I_1\;. 
\end{eqnarray}
Finally, by partial integration, using (\ref{a01}) and taking 
derivatives with respect to $\ql ,\qlp$ we find the
result given by (\ref{gf28}).

\end{appendix}

\end{document}